\documentclass[apj]{emulateapj}

\shorttitle{Electrons and Alkalis in Exoplanets}
\shortauthors{Lavvas et al.}

 \usepackage{color}
\newcommand{\clr}{black}

\begin{document}


\title{Electron densities and alkali atoms in exoplanet atmospheres}


\author{P. Lavvas\altaffilmark{1}, T. Koskinen\altaffilmark{2}, and R.V. Yelle\altaffilmark{2}}
\altaffiltext{1}{GSMA, Universit\'e de Reims Champagne Ardenne, CNRS UMR 7331, France}
\altaffiltext{2}{Lunar and Planetary Laboratory, University of Arizona, Tucson, AZ 85719, USA}
\email{panayotis.lavvas@univ-reims.fr}



\begin{abstract}
We describe a detailed study on the properties of alkali atoms in extrasolar giant planets, and specifically focus on their role in generating the atmospheric free electron densities, as well as their impact on the transit depth observations. We focus our study on the case of HD 209458 b, and we show that photoionization produces a large electron density in the middle atmosphere that is about two orders of magnitude larger than the density anticipated from thermal ionization. Our purely photochemical calculations though result {\color{\clr}in} a much larger transit depth for K than observed for this planet. This result does not change even if the roles of molecular chemistry and excited state chemistry are considered for the alkali atoms. {\color{\clr}In contrast}, the model results for the case of exoplanet XO-2 b are in good agreement with the available observations. Given these results we discuss other possible scenarios, such as changes in the elemental abundances, changes in the temperature profiles, and the possible presence of clouds, which could potentially explain the observed HD 209458 b alkali properties. {\color{\clr}We find that most of these scenarios can not explain the observations, with the exception of a heterogeneous source (i.e.~clouds or aerosols) under specific conditions, but we also note the discrepancies among the available observations.}
\end{abstract}





\section{Introduction}

Alkali atoms and free electrons in planetary atmospheres are strongly related as the low ionization potentials of the former can induce significant populations of the latter. Alkali atoms are expected to be abundant in hot atmospheres where their condensation or loss to other molecular structures will be minimal. In this sense, the hot atmospheres of close-in exoplanets are excellent candidates for the investigation of the alkali-electron relationship, and its subsequent ramifications on the atmospheric electro-dynamics. 

Although alkali atoms are now observed in many different exoplanets, we focus our study on the case of HD 209458 b, due to the extended number of observational constraints available for this planet. Observations with the Hubble Space Telescope (HST) at visible wavelengths provided the first detection of an exoplanet atmosphere through the detection of sodium in HD 209458 b \citep{Brown01, Charbonneau02}. Subsequent observations with the same instrument \citep{Ballester07,Knutson07}, reproduced the first detection and verified that the Na absorption was smaller than anticipated theoretically {\color{\clr}\citep{Seager00,Hubbard01,Brown01b}}, while potassium and lithium were not detected. Ground based observations in the visible also detected Na in the atmosphere of HD 209458 b, and verified the lack of K in the measured spectra \citep{Snellen08,Jensen11}. 

Based on these measurements, many subsequent studies have focused on their re-analysis and interpretation through different scenarios. \cite{Barman02} suggested that non-LTE effects could be responsible for the decreased Na transit depth observed, while \cite{Fortney03}, following on the initial scenarios suggested by \cite{Charbonneau02}, investigated the role of ionization and potential cloud formation, and suggested that both processes are important for the absorption in the Na D line core. \cite{Sing08a} re-analyzed the HST observations at a higher resolution and suggested that a local temperature minimum must exist close to the mbar pressure level of HD 209458 b, which would allow Na to condense and reduce its transit depth signature. {\color{\clr}\cite{VidalMadjar11} used the differential transit depth measurements reported by \cite{Sing08a} to derive a vertical thermal profile of HD 209458 b's atmosphere, assuming a constant with altitude Na abundance.} The lack of K in the absorbed spectra, however, has not been investigated in detail before. The commonly anticipated scenarios to explain this lack are a strong photochemical loss (due to the lower ionization of K relative to Na), the screening of the K signature by a cloud/haze, or the lower than solar abundance of this element in the atmosphere of HD 209458 b \citep{Desert08}.
 
In the meanwhile, measurements in other regions of the electromagnetic spectrum, as well as theoretical studies provide more information about the atmospheric properties of HD 209458 b. Transit observations with HST at shorter wavelengths indicate the presence of HI, CII, OI, and SiIII in the thermosphere of this exoplanet \citep{VidalMadjar03, VidalMadjar04, BenJaffel10, Linsky10}, which impose important constraints on both the temperature of the thermosphere, as well as the efficiency of condensation of heavy elements in the deeper atmosphere \citep{Koskinen10,Koskinen13a,Koskinen13b}. Observations at IR wavelengths during primary \citep{Beaulieu10, Deming13} and secondary \citep{Knutson08,Swain09,Crossfield12} eclipse reveal the presence of  H$_2$O, CH$_4$, and CO$_2$ in the deep atmosphere of HD 209458 b, and inform us on the temperature conditions of the day side disk and of the terminator. These observations have helped to constrain general circulation models \citep{Showman09} and photochemical models \citep{Moses11}. Such simulations are important because they provide information for the role of dynamics in the redistribution of energy deposited on the day side to the night side, the role of atmospheric mixing, and the possible abundances of the important chemical species that are suggested by the observations.  

In light of the new information based on a combination of observations and models, we use a photochemical model to investigate the role of Na and K in the atmosphere of HD 209458 b. We focus on the relative abundance of Na and K in order to address the surprising lack of a K signature in the observations. Given the detection of K in other exoplanet atmospheres such as XO-2 b \citep{Sing10}, this study is a timely introduction to a comparative investigation into the role of alkali atoms in different exoplanet atmospheres. We first discuss the impact of photochemistry on the vertical profiles of these elements and the produced electron densities, and then proceed to a more thorough study of the role of the excited state chemistry of these elements. Finally, based on the conclusions we derive from our calculations we discuss all the previously suggested scenarios for the interpretation of the observations. The impact of the free electron densities derived here on the atmospheric electro-dynamics is discussed in the accompanying paper \citep{Koskinen14}.

\begin{figure}
\centering
\includegraphics[scale=0.5]{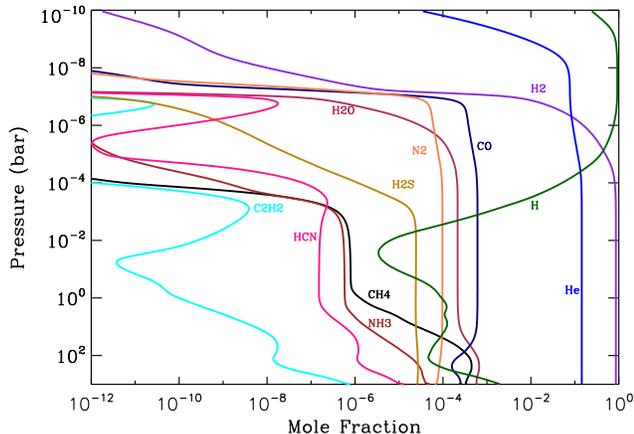}
\caption{Major chemical species for HD209458b calculated by our model.}\label{chem_HD209}
\end{figure}

\section{Model description}

We use a 1D photochemical model that extends from 10$^3$ to 10$^{-10}$ bar. Our model is based on a previous version applied to Titan's atmosphere \citep{Lavvas08}, but has been extended to include forward and reverse reactions \citep[with thermochemical parameters from][]{Burcat05}, and is therefore able to kinetically simulate thermochemical processes for species containing H, C, N, O, S and Si. The model solves the continuity equation for each species taking into consideration photochemical and thermochemical sources and sinks, and the transport of the species due to molecular diffusion and atmospheric mixing (the latter based on an eddy diffusion profile). In total we include $\sim$130 species involved in $\sim$1600 reactions. In this study we focus on the processes of ionization and chemistry of alkali species, which have important implications for the interpretation of the available observations. We avoid a full description of the basic H/C/N/O/S chemical networks, which have already been well described in previous studies \citep[e.g.][]{Zahnle09,Moses11,Venot12}, and focus our study on Na and K chemistry. For the case of HD 209458 b our model results for the main atmospheric compounds (Fig.~\ref{chem_HD209}) are similar to those of previous studies.  

\begin{figure}
\centering
\includegraphics[scale=0.52]{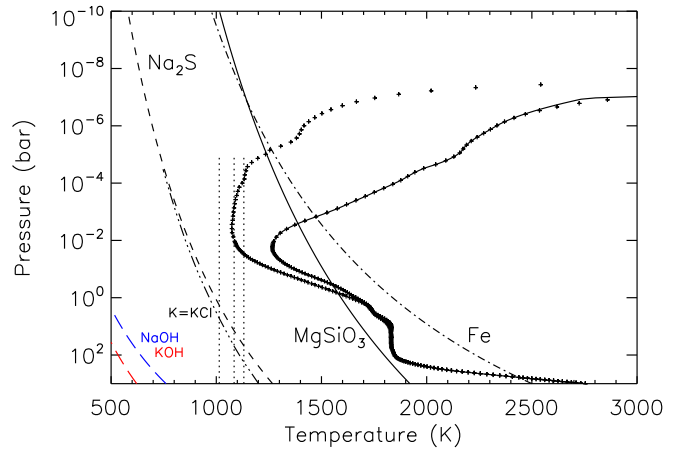}
\caption{Temperature-pressure profiles assumed for day-time (crosses with solid line) and terminator average (crosses) conditions of HD209458b based on \cite{Showman09}. Condensation curves for Fe (dash-dotted), MgSiO$_3$ (solid), and Na$_2$S (dashed) are also shown. The K=KCl curve marks the transition from atomic K to molecular KCl under thermochemical equilibrium, while the colored dashed lines show the condensation curves of NaOH and KOH. The vertical dotted lines present skin temperatures for HD 209458 b (1085 K for bond albedo of A=30$\%$ and 1130 K for A=17$\%$) and XO-2 b (1014 K for A=30$\%$).   }\label{temps}
\end{figure}

For the atmospheric structure of HD 209458 b we assume the vertical temperature - pressure profiles suggested by general circulation models \citep{Showman09} for the disk and terminator conditions (Fig.~\ref{temps}). Above 10$^{-6}$ bar we have merged these profiles with the thermospheric temperature profile calculated by \cite{Koskinen13b}. For the atmospheric mixing we consider the eddy diffusion profile derived from GCM for the case of HD 209458 b \citep{Moses11}. {\color{\clr} \cite{Parmentier13} demonstrate that the eddy profile used by \cite{Moses11} is likely too large for HD 209458 b, therefore we discuss below the effects of different mixing profiles.} At the lower boundary we assume solar elemental abundances and thermochemical equilibrium \citep{Lodders10}, while at the upper boundary we allow species with mass less than 3 amu to escape with a velocity provided by escape models \citep{Koskinen13a,Koskinen13b}. The effects of hydrodynamic escape on the heavier species are not included here as we focus on the lower atmosphere.

We use ionization cross sections from \cite{Verner96} for Na, and from \cite{Sandner81} and \cite{Zatsarinny10} for K. For Na we use recombination rates from \cite{Verner96} for the radiative process and from \cite{Su01} for the 3-body process. The latter study provides only the low pressure limit rate and we assumed a high pressure limit of 10$^{-7}$ cm$^3$s$^{-1}$. For K the radiative recombination rate \citep{Landini91} is very close to the Na radiative rate, and as we did not find any measurements for the 3-body recombination of K, we assumed the same recombination rates as for Na. Thermal ionization rates were calculated on the basis of microscopic reversibility based on the assumed rates of 3-body recombination. 

\begin{figure}
\centering
\includegraphics[scale=0.5]{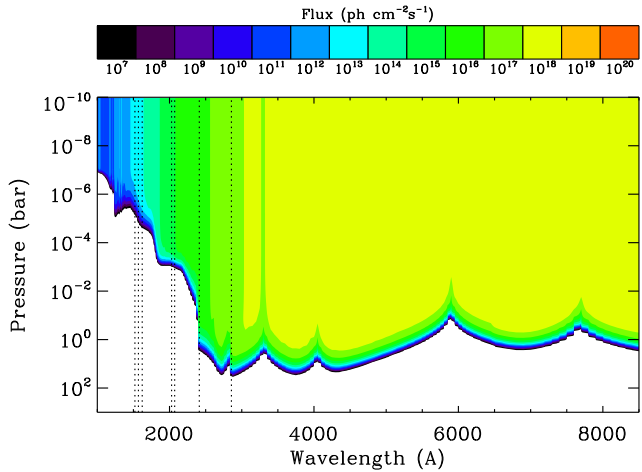}
\caption{Stellar flux penetration in the atmosphere of HD209458b for different wavelengths. The vertical dotted lines show the ionization limits of the heavy atoms included in the calculations (see Table~\ref{elements}), while absorption by Na and K resonant lines dominates at longer wavelengths.}\label{RadField}
\end{figure}

For the stellar spectrum of HD209458 we assume a solar-average spectrum that is representative of G type stars like HD 209458. Atmospheric attenuation includes opacities from all optically significant species included in the model, and Rayleigh scattering by H$_2$. Fig.~\ref{RadField} presents the attenuated radiation field at different pressure regions of the atmosphere at UV and visible wavelengths. At the ionization limits of Na and K (see Table~\ref{elements} and vertical lines in Fig.~\ref{RadField}) significant photon fluxes persist to the deep atmosphere and can induce large ionization rates as we describe below.

\begin{table} 
\centering
\caption{Properties of atoms contributing to ionization in the lower atmosphere. Abundances are given in terms of a mixing ratio to H$_2$ considering a He mixing ratio of 0.144. For the special case where we consider the presence of other heavy atoms than Na and K (section 3.3), at the lower boundary aluminum is in the form of AlOH and AlH, while silicon forms SiO and SiH$_4$. All other elements are dominantly in their atomic form.{\color{\clr} Resonance lines reported correspond to dominant transitions in the region between 300 and 800 nm. For Fe see Fig.~\ref{HD209_trans_metals}. For doublet lines we give the mean wavelength. }}\label{elements}
\begin{tabular}{ccccrrc}
\hline
Z   & Element & Atomic  & Mixing & \multicolumn{2}{c}{Ion. Limit} & $\lambda_{Res}$ \\ 
	& 		 &Mass	&	Ratio	& eV 			& 	$\rm\AA$ & nm		\\ 
\hline
		11 	& Na 	& 23			&3.4$\times$10$^{-6}$& 5.139 	& 2412.3 & 589.3 \\ 
		12 	& Mg 	& 24.3		&5.9$\times$10$^{-5}$& 7.646 	& 1621.6 & 457.1\\ 
		13 	& Al		& 27		 	&5.1$\times$10$^{-6}$& 5.986	& 2071.2 & 308.2, 394.4\\ 
		14 	& Si		& 28.1		&5.7$\times$10$^{-5}$& 8.152	& 1520.9 &-\\ 
		19 	& K		& 39.1	 	&2.3$\times$10$^{-7}$& 4.340	& 2856.8 & 768.2\\ 
		20 	& Ca		& 40.1 		&3.7$\times$10$^{-6}$& 6.113 	& 2028.2 & 422.7\\ 
		26 	& Fe 		& 55.8 		&4.8$\times$10$^{-5}$& 7.902 	& 1569.0 & multiple\\ 
\hline
\end{tabular}\\
\end{table}

\section{Results}
We discuss here the ion and electron densities, resulting from the photoionization of the alkali atoms, and compare our calculated transit depth spectrum, based on the calculated mixing ratio profiles, with the available observations for HD 209458 b. We also discuss possible contributions from the ionization of other atomic compounds. 

\subsection{Ions $\&$ Electrons}

Due to their different ionization potentials, ionization of each atom is initiated at a different pressure level in the atmosphere. In the upper atmosphere we consider ionization of H, He, C, N, and O, the ion-chemistry of which we have already discussed in \cite{Koskinen13a}. These elements have large ionization potentials and they do not significantly affect the ionization of the alkali atoms deeper in the atmosphere. {\color{\clr}In order to demonstrate the dominance of the alkali atoms in the ionization of the lower atmosphere, here we only discuss the effects of their photoionization and electron recombination. Additional processes that could affect the profiles of the alkali atoms are discussed in section 4.}

\begin{figure}[!t]
\centering
\includegraphics[scale=0.5]{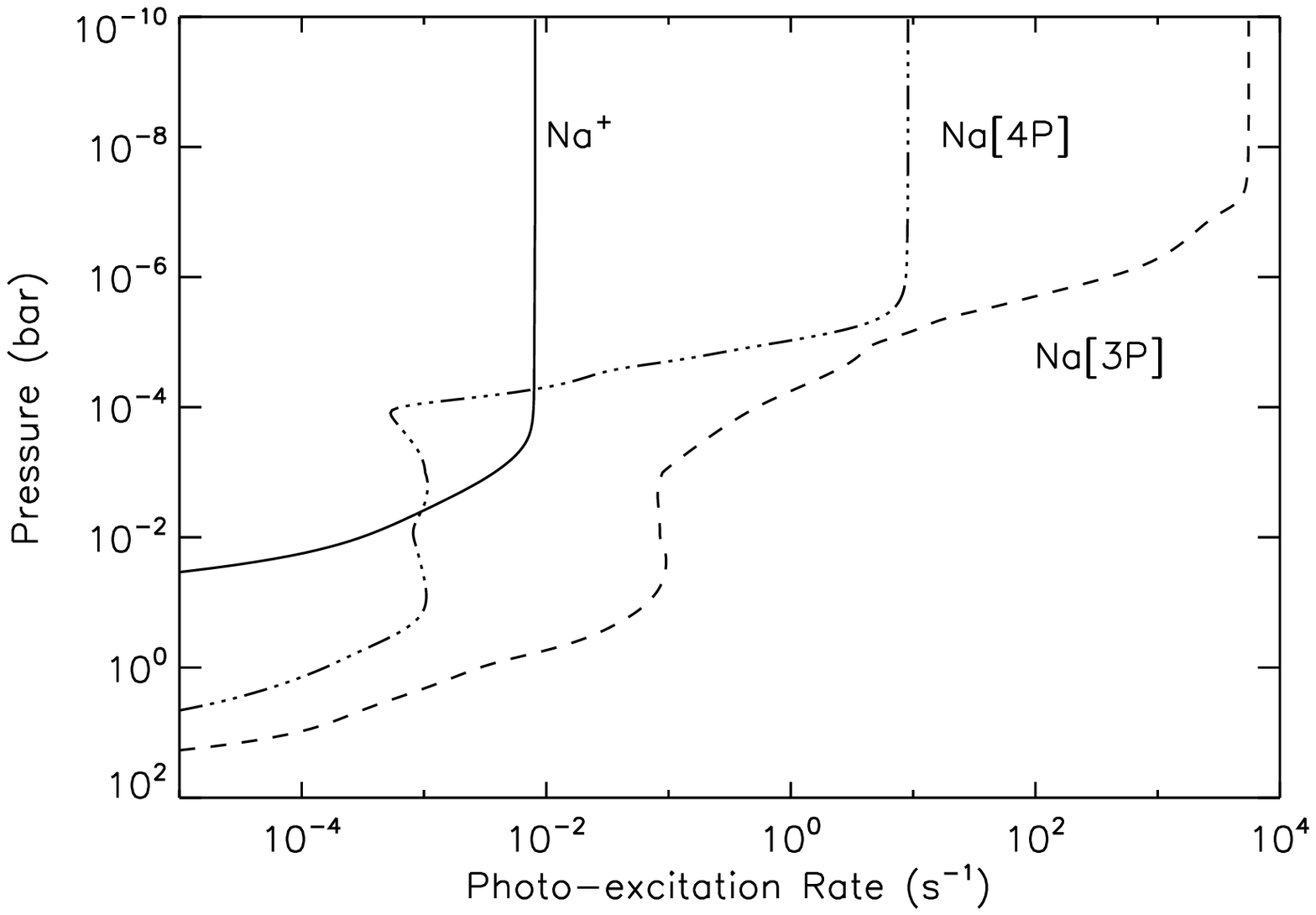}
\includegraphics[scale=0.5]{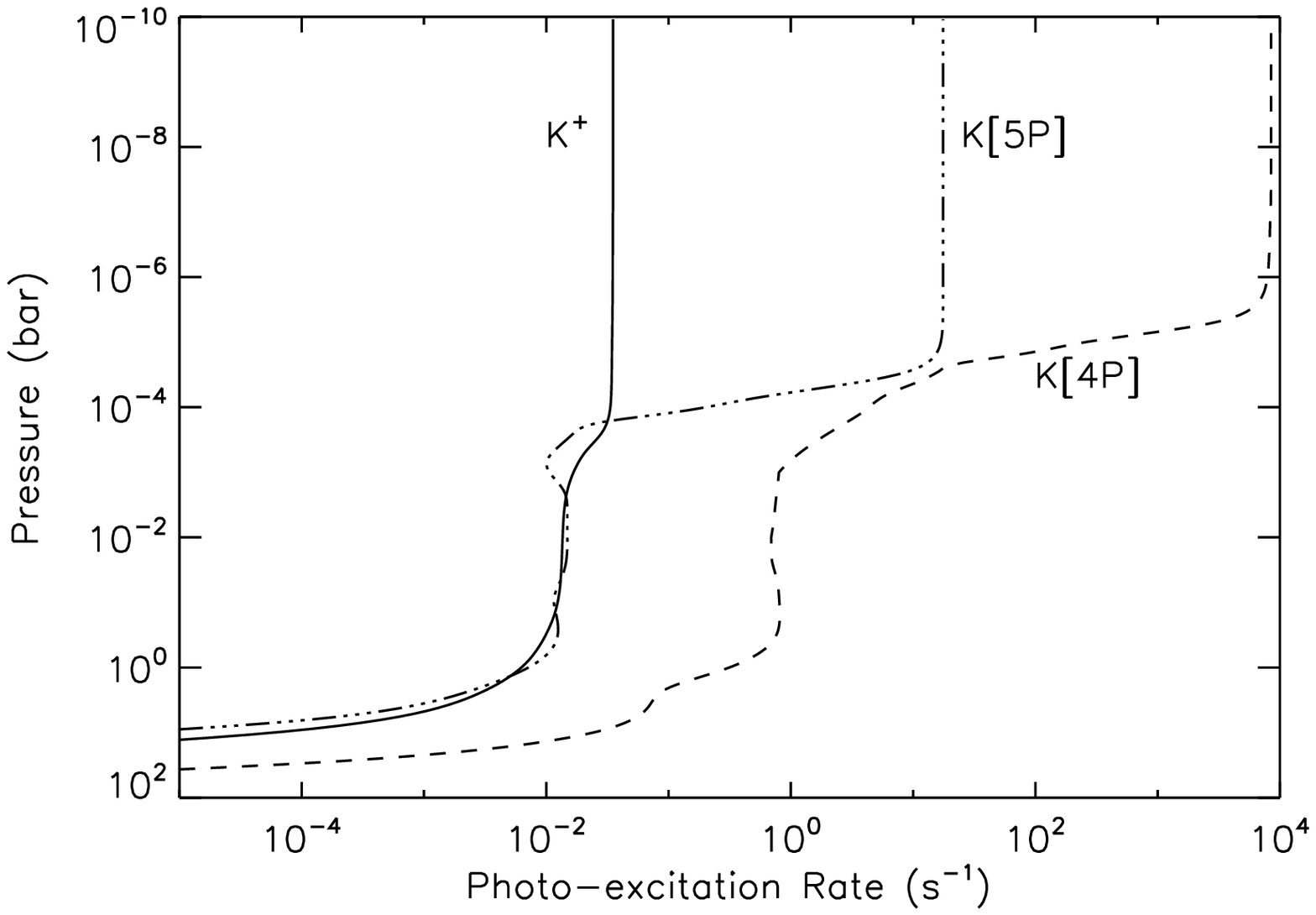}
\caption{Photoionization rates (solid lines) of Na (top) and K (bottom). For comparison the corresponding rates for photo-excitation of different resonant states of are shown.}\label{jrates}
\end{figure}

\begin{figure}[!t]
\centering
\includegraphics[scale=0.5]{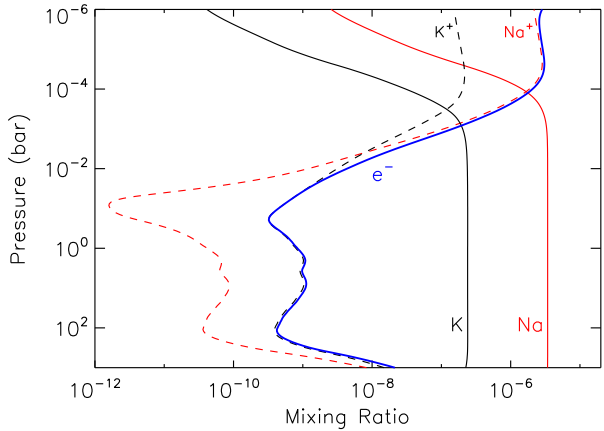}
\caption{Vertical profiles for the mixing ratios of Na (red) and K (black) from our calculations. Solid lines show the neutral compound and dashed lines the corresponding ion. The total electron mixing ratio profile is presented by the blue line. }\label{Ions_mr}
\end{figure}

Potassium has the most extended photoionization cross section reaching to $\sim$2900 $\rm\AA$. Photons at these wavelengths are not substantially absorbed by the other atmospheric gaseous compounds resulting in ionization of K to pressures greater than 1 bar (Fig.~\ref{jrates}). For Na, photons available for ionization with $\lambda \le$ 2412 $\rm\AA$ are also absorbed by other gaseous species, such as H$_2$O and H$_2$S (as identified also by \cite{Fortney03}) and photoionization is efficient down to $\sim$10$^{-2}$ bar. At low pressures Na$^+$ and K$^+$ recombine radiatively with electrons. This process is slow, resulting in large electron densities. At deeper pressures, greater than 10$^{-3}$ bar, Na$^+$ and K$^+$ are dominantly lost through 3 body recombination, which is much more rapid. Hence, most of the ionized alkali atoms return to their neutral state and the electron densities decrease (Fig.~\ref{Ions_mr}). At even higher pressures, larger than 0.1 bar, thermal ionization becomes efficient and the electron densities start to increase again. The 3-body recombination also becomes more efficient with increased pressure though, forcing the majority of the alkali atoms to remain in their neutral state.

\begin{figure}[!t]
\centering
\includegraphics[scale=0.5]{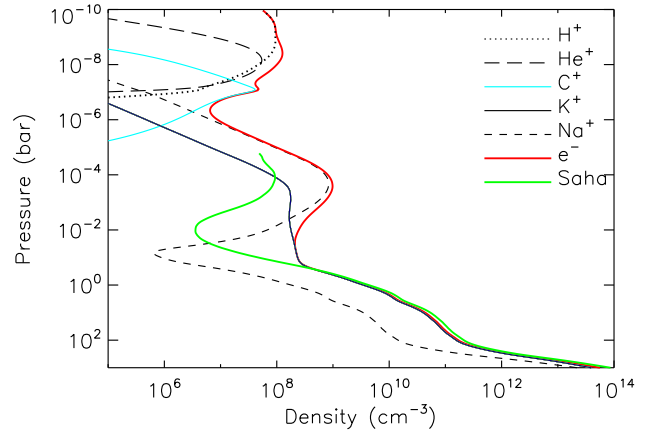}
\caption{Calculated densities of different ions and the resulting electron density (solid red line) assuming charge balance, in the atmosphere of HD 209458 b. The green line presents the electron density assuming thermal ionization only. }\label{Ion_den}
\end{figure}

The variable ionization regions of these atomic elements have their signature on the resulting electron densities (Fig.~\ref{Ion_den}). For pressures lower than 10$^{-8}$ bar electrons are produced from the ionization of atomic hydrogen, with secondary contributions by He and C photoionization. At higher pressures the majority of the electrons is produced by the photo-ionization of Na and K, with the former dominating over the latter up to pressures of $\sim$10$^{-2}$ bar. Below the 1 bar level photo-ionization rates are small and most of the electrons are produced by thermal ionization of K. Although thermal ionization produces large electron densities below the 1 bar level, it is important to note that photo-ionization dominates at lower pressures. The evaluation of the thermal electron population based on the Saha equation considering all atomic elements used in the kinetic model shows that the thermal electron density at 10$^{-2}$ bar should be $\sim$4$\times$10$^{6}$ cm$^{-3}$ (green line in Fig.~\ref{Ion_den}). The photon-induced electron density at the same level is almost 100 times larger. Such high electron densities so deep in the atmosphere are not common among the giant planets of our solar system, and for HD 209458 b result from the combination of strong stellar fluxes for this close-in exoplanet, and  high atmospheric temperatures that allow Na and K to remain in atomic form. Thus, we can conclude that photo-ionization in the atmospheres of exoplanets similar to HD 209458 b will produce large electron densities in the middle atmosphere between 10 $\mu$bar and 1 bar.

\subsection{Transit depths}

\begin{figure*}[!t]
\centering
\includegraphics[scale=0.75]{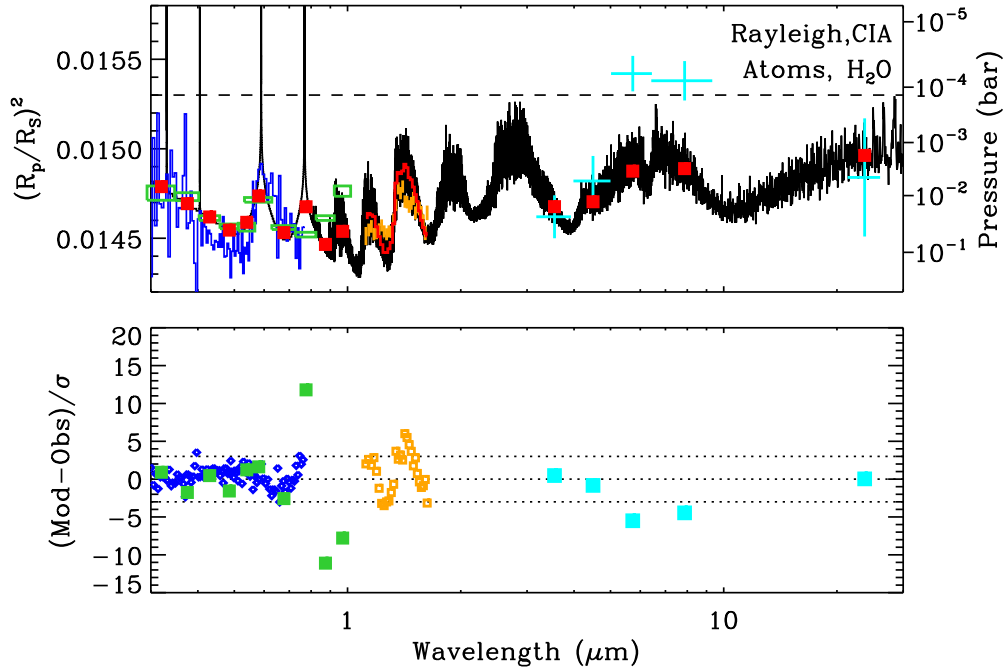}
\caption{Transit depth calculation in the atmosphere of HD 209458 b. In the top panel, the black line is the transit depth based on our calculations including all atoms considered, Rayleigh scattering by H$_2$, collision induced absorption, and H$_2$O. The green boxes present the results from \cite{Deming13} from the re-analysis of the HST/STIS observations of \cite{Knutson07}, while the blue histogram presents the HST/STIS observations from \cite{Sing08a}. The orange lines present the HST/WFC observations from \cite{Deming13}. The cyan cross are the Spitzer transit depth measurements from IRAC \citep{Beaulieu10} and MIPS \citep{Crossfield12}. The red points and line present our calculations mapped to the resolution of the observations. The dashed line presents the required cloud transit depth required to match the differential transit observations. In the lower panel we present the residuals between model and observations in terms of the 1 sigma level of uncertainty in the observations. The blue points correspond to the comparison with the \cite{Sing08a} data, the green and orange squares to the \cite{Deming13} data from STIS and WFC data, respectively, and the cyan squares to the Spitzer observations.}\label{HD209_trans}
\end{figure*}

We now turn our attention to the comparison of our model results with the available observations of alkali metals in HD 209458 b. Our opacities for the transit depth calculation include H$_2$ Rayleigh scattering, absorption by Na and K, collision induced absorption by H$_2$-H$_2$ and H$_2$-He collisions, and molecular absorption by H$_2$O (Fig.~\ref{HD209_trans}). We use the atomic line properties from the NIST database and we calculate the line profiles for the different pressure-temperature conditions of the model atmosphere assuming Voigt line profiles. For the alkali atoms we use the line parameters and broadening coefficients from \cite{Iro05}. We are aware of the divergence of the alkali line wings from the Voigt description at the far wings \citep{Allard03}, but the conclusions we draw below are not affected by this issue. For the temperature profile we assume the average terminator profile (see Fig.~\ref{temps}), while for the chemical abundances we use the densities derived by the day side conditions since we expect that the strong zonal transport suggested by the GCMs will efficiently redistribute the abundances from the disk to the terminator \citep{Showman09}. We discuss further below possible implications for the differences between the dawn and dusk terminator.

\begin{figure}[!t]
\centering
\includegraphics[scale=0.5]{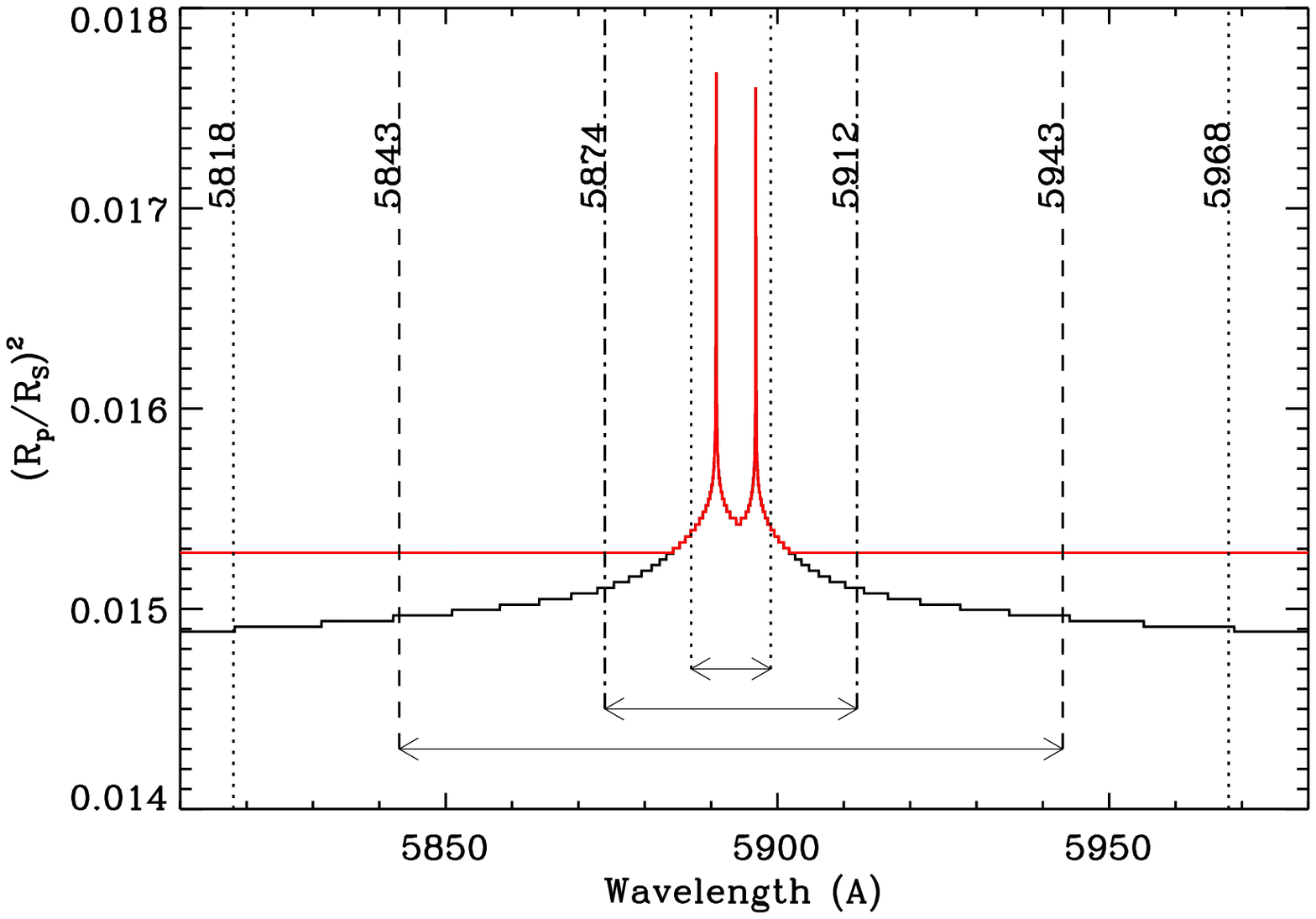}
\includegraphics[scale=0.5]{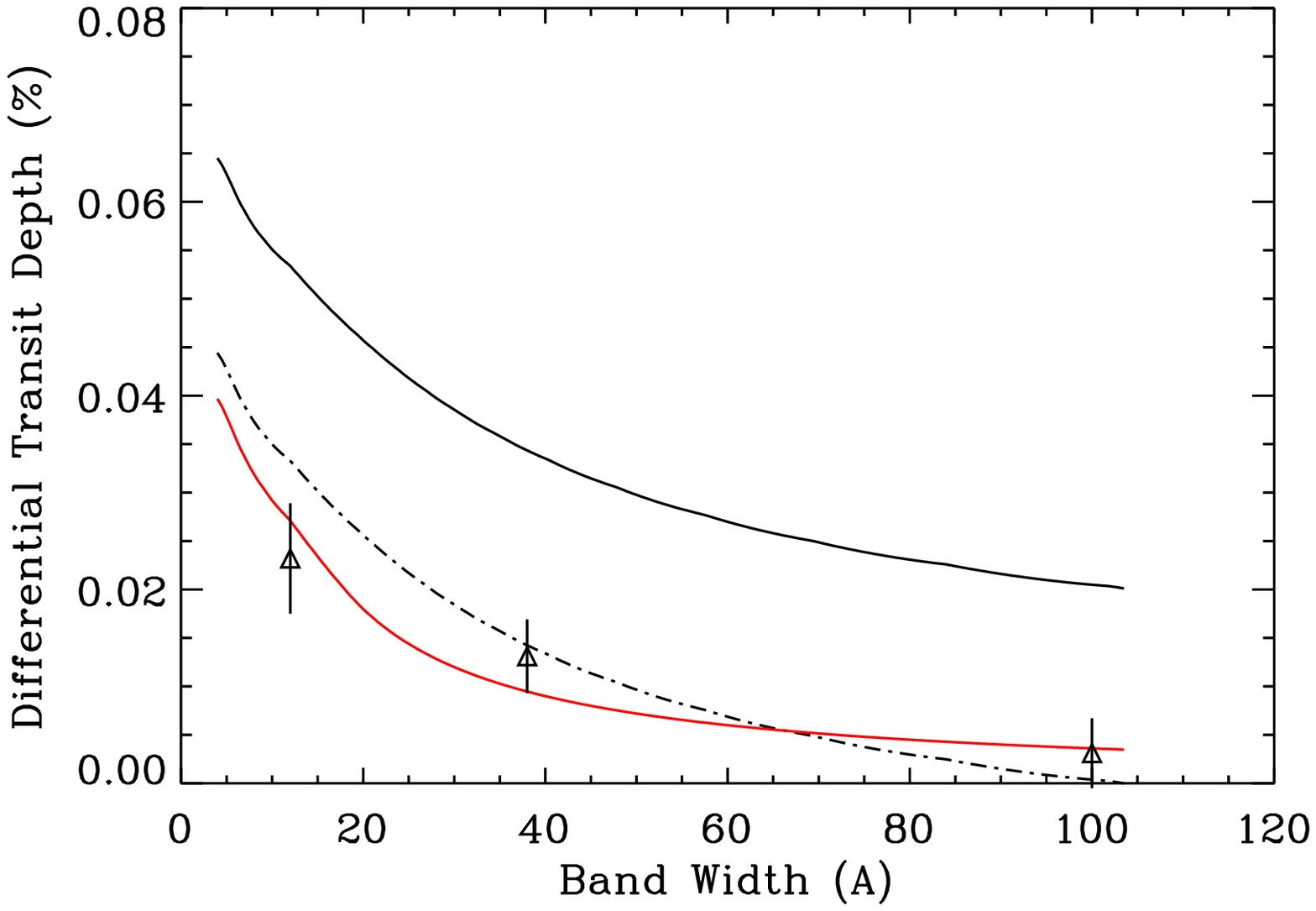}
\caption{Top: Transit depth spectrum around the Na line core. The black line corresponds to the clear atmosphere case and the red line to the case with a cloud deck. The vertical broken lines designate the bands used by \cite{Charbonneau02} for the investigation of the differential transit depth assuming band widths of 12, 38, and 100 \AA (arrows). Bottom: Differential transit depth calculation performed by assuming different band widths around the Na core and comparing with the reference region bound between 5818 and 5968 \AA. The solid line is the clear atmosphere case assuming the Na profile from our photochemical model, while the dash-dotted line corresponds to the clear atmosphere case but homogeneously shifted downwards by $\sim$0.02$\%$. The red line corresponds to the differential transit depth assuming a cloud deck with a transit depth of $\sim$0.0153 around the sodium core. The symbols with the associated error bars are the differential transit depth values from the \cite{Charbonneau02} HST/STIS observations.}\label{diff_trans}
\end{figure}


The transit depth at the sodium line appears to be consistent with the sodium wings observed by \cite{Sing08a} in the medium resolution STIS observations (Fig.~\ref{HD209_trans}). The analysis of the same observations by \cite{Knutson07} and \cite{Deming13} at larger spectral bins suggests the presence of a less broad Na wing, but our model is consistent with these observations as well, given the larger uncertainty in the mid-resolution observations. The presence though or not of a broad wing at the sodium line has important implications for the interpretation of the observations and specifically the question of whether the atmosphere is clear or not. A broad atomic line wing implies a clear atmosphere, while a shallow atomic core would indicate the presence of clouds (or haze) at the probed altitudes. 


A closer look at the Na line double core (Fig.~\ref{diff_trans}) shows that our model generates a differential transit depth that is larger than the original observations by \cite{Charbonneau02} and the subsequent observations by \cite{Sing08a}. The very low differential transit depth at the sodium line core was originally interpreted as an indication of a cloud deck at the probed pressure levels around the sodium core that reduced the contrast between the line center and the wings \citep{Charbonneau02}. We find that such an interpretation is not consistent with the subsequent observations of Rayleigh scattering at shorter wavelengths, and in general with the low resolution spectra. This argument becomes obvious if we assume a spectrally flat cloud deck around the Na core that would be consistent with the observed differential transit depth values. Our calculations suggest that this would occur for a cloud transit depth close to 0.00153 (see Fig.~\ref{diff_trans}), which results in a differential transit depth that is consistent with both the magnitude and the spectral dependence of the observations. However, such a cloud would be inconsistent with the observed transit depth at shorter wavelengths since it has a larger absorption than the strongest absorption observed at 0.3 $\mu$m (see dashed line in Fig.~\ref{HD209_trans}). In other words the cloud properties implied by the high resolution differential transit depth are not consistent with the Na wing detection and the observed slope at shorter wavelengths\footnote{In reality the cloud spectral signature will not be flat over the whole spectral range of the observations, but for small particles (as expected at the pressure regions of interest here) the absorption would increase towards shorter wavelengths and decrease towards longer wavelengths around the sodium core. Such a spectral behavior would make the cloud scenario even less consistent with the differential transit depth observations, relative to the spectrally flat case.}. We further discuss the implications of this result below.


At longer wavelengths, absorption by H$_2$O alone provides a consistent picture with the HST observations by \cite{Deming13} between 1.1 and 1.6 microns, and the Spitzer observations at longer wavelengths by \cite{Beaulieu10} and \cite{Crossfield12}, although the model seems to lack some opacity at 5.8 and 8 microns. 

The most striking feature of the  comparison of the model with the observations is the large transit depth at the potassium line in the model relative to practically no detection in the observations \citep{Knutson07,Sing08a,Jensen11}. It is commonly anticipated that the lack of potassium could be explained by the rapid photo-ionization of this element due to its very low ionization potential. Our transit depth calculations imply that for the assumed solar elemental abundances, the K profile would have to be depleted to very deep levels in the atmosphere (close to 10 mbar) in order to be consistent with the observations. Our photochemical calculations presented above demonstrate that this is not likely. Before considering other options to explain this discrepancy between observations and modeling, we decided to further investigate the alkali chemistry in order to verify that other photochemical processes can not help reduce the abundances of Na and K. These studies are described in section 4.

\subsection{Other atoms}

We have also considered the possibility of other atoms in the atmospheres of hot extrasolar giant planets. For example, Mg, Si, Ca,  Al, and Fe combine a low ionization potential with a high elemental abundance relative to other heavy atoms (Table~\ref{elements}) and could contribute to the electron densities. Most of these heavy atoms should condense in the low atmosphere for the temperature conditions assumed (see Fig.~\ref{temps}), in which case they would not significantly contribute to the ionosphere. The observations of HD 209458 b seem to favor an atmosphere in which Si, at least, survives in the upper atmosphere. Specifically, the signature of SiIII in the transit depth observations \citep{Linsky10}, implies a large abundance of atomic silicon in the thermosphere, which should not be possible if condensates such as forsterite or enstatite form in the lower atmosphere \citep{Koskinen13a,Koskinen13b}. Based on these observations and interpretation, we performed calculations including the above atoms in order to evaluate their potential impact on the electron densities and the transit depth. 

We used ionization cross sections for all new atoms from \cite{Verner96}. We could not find recombination rates for all elements considered, thus certain estimations had to be made based on the electronic structure of each element. For Mg and Si we used recombination rates from \cite{Aldrovandi73} and for Ca we assumed the same rate as for Mg. For Al we estimated the recombination rate with that of SiIII from \cite{Aldrovandi73}, while for Fe we used the recombination rate from \cite{Nahar97}. Three body rates were estimated based on the rate for sodium. Based on the NASA thermochemistry model \citep{Burcat05}, for the atmospheric conditions of HD209458b, the most abundant form of the heavy elements is the atomic, with the exception of Si and Al. Aluminum at the lower boundary is dominantly in the form of AlOH and AlH, which we follow kinetically in the model through the reactions: 
\begin{center}
Al + H$_2$O $\leftrightarrow$ AlOH + H  \\
Al + H + M $\leftrightarrow$ AlH + M 
\end{center}
with rates for the forward reactions taken from \cite{McClean93, Sharipov11}, and \cite{Swihart03}, respectively. Silicon, like aluminum, is present in molecular compounds at the lower boundary with SiO the most abundant species. For this case though there is a large number of other silicon species that can be formed. In our calculations we have followed kinetically the transition between SiO and SiH$_4$ through different intermediate species, as well as the interaction of silicon with other elements to form SiS, SiN, and SiC. {\color{\clr}The SiO-SiH$_4$ equilibrium is described kinetically with a scheme similar to the CO-CH$_4$ equilibrium \citep{Moses11}, but modified by the thermochemical parameters of the silicon-based species. For the most part of the atmosphere SiO and SiH$_4$ remain close to their equilibrium abundances.} The details of these calculations are beyond the scope of the current study and will be presented in a future study.

\begin{figure}[!t]
\centering
\includegraphics[scale=0.5]{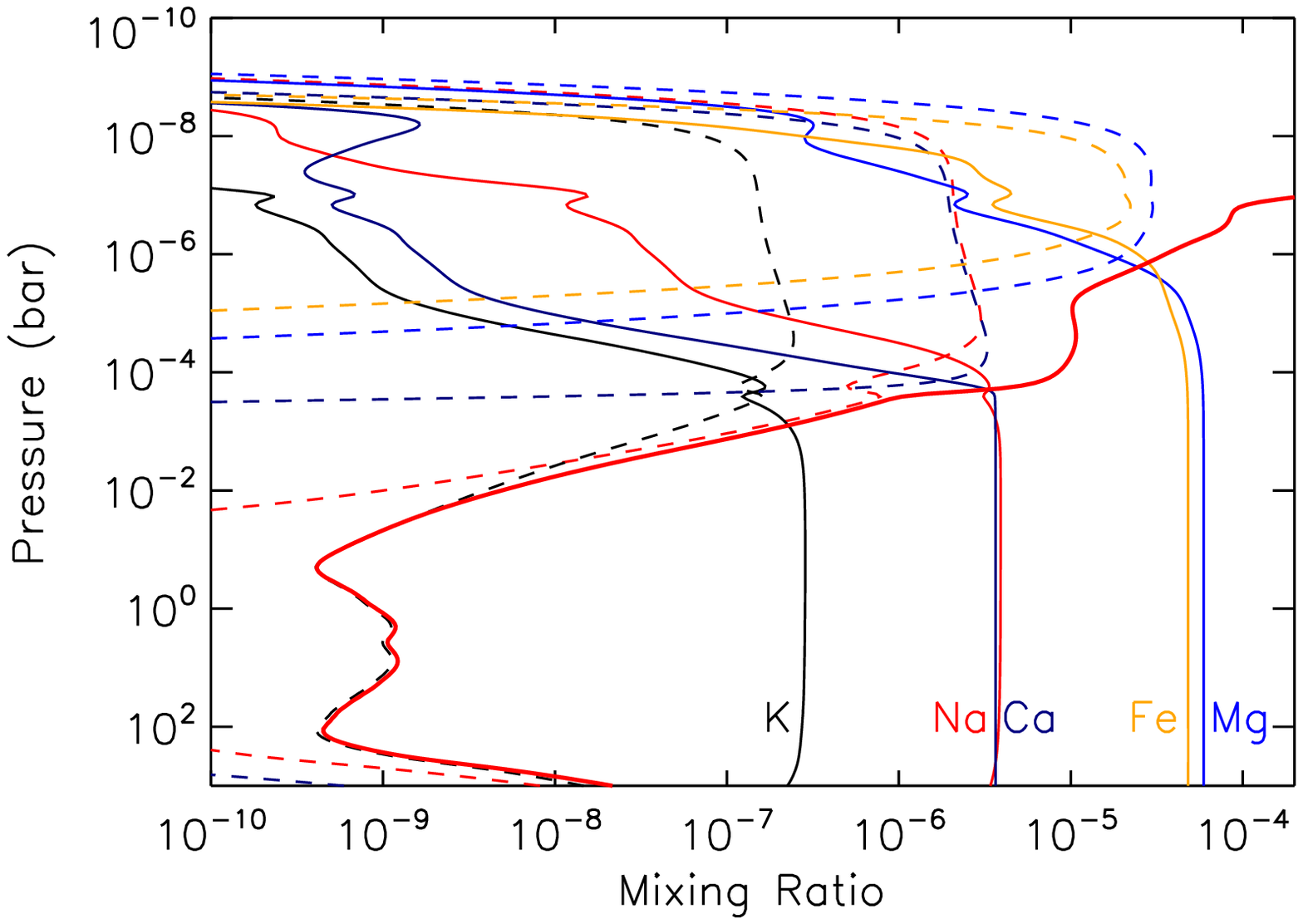}
\includegraphics[scale=0.5]{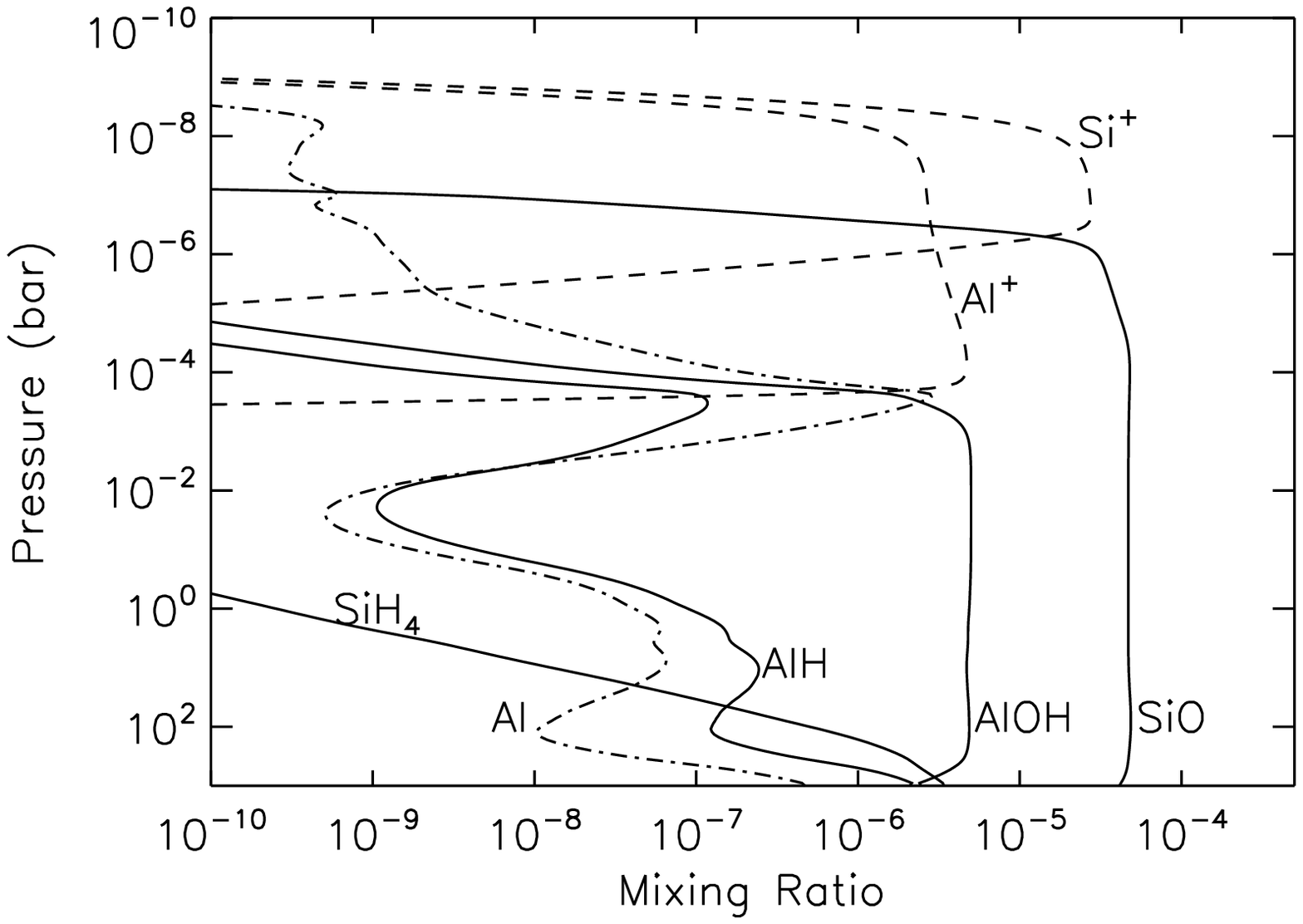}
\caption{Vertical profiles for the mixing ratios of atoms and their corresponding ions included in the calculations. The top panel presents the profiles for elements that are present in their atomic form and bottom panel presents the case of aluminum and silicon species that are present in molecular form at the lower boundary. Solid lines show the neutral compound and dashed lines the corresponding ion. For clarity the Al profile is presented with a dash-dotted line.}\label{Ions_mrB}
\end{figure}


According to our calculations, the two aluminum molecular compounds have similar abundances at the lower boundary but diverge at lower pressures with AlOH becoming the most abundant species close to 10$^2$ bar (Fig.~\ref{Ions_mrB}). At pressures below 10$^{-2}$ bar, reaction with atomic hydrogen gradually returns aluminum to its atomic form before becoming mainly ionized at pressures lower than 10$^{-3}$ bar. Calcium has a similar ionization pressure region since its ionization limit is close to that of aluminum. The remaining elements Mg, Fe and Si have all significantly higher ionization potentials compared to the previous elements, and their corresponding ions become important only above 10$^{-6}$ bar. Mg, Fe, and Si photo-ionization contributes to the local increase of the electron density close to 10$^{-6}$ bar, while the signature of Ca and Al photo-ionization is clear close to 10$^{-4}$ bar. The increased electron abundance also affects the Na and K profiles close to 10$^{-4}$ due to the enhanced recombination rates. 

\begin{figure}
\centering
\includegraphics[scale=0.52]{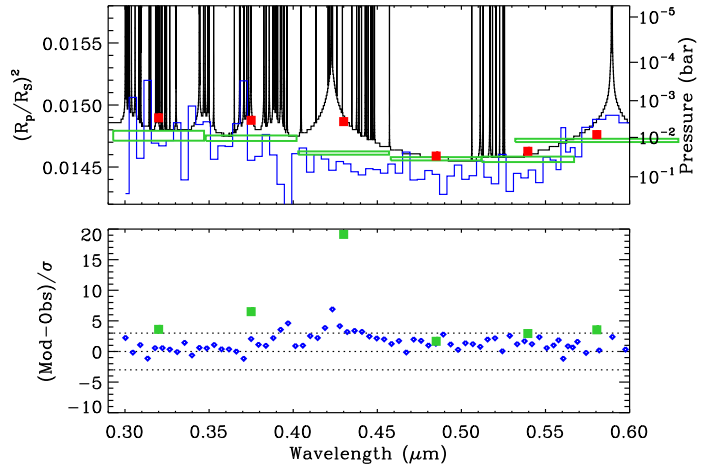}
\caption{Same as Fig.~\ref{HD209_trans} but with the inclusion of the extra atomic abundances calculated by the model.}\label{HD209_trans_metals}
\end{figure}

Inclusion of these new atomic profiles to the transit depth calculations produces numerous narrow absorption lines at short wavelengths \citep[see also][]{Barman07}. Although most of these lines are too narrow to be detectable with the current instrumentation, some of them, originating mainly from {\color{\clr}Ca ($\sim$423 nm) and Fe (all other broad absorption features between 0.3 and 0.4 $\mu$m)}, are wide enough to be observed. The medium resolution HST/STIS observations \citep{Sing08a} demonstrate some structure in this part of the spectrum, but they suffer from large uncertainties that do not allow for a clear assignment of absorption signatures. On the other hand, the low resolution analysis of the same observations \citep{Knutson07,Deming13} do not indicate the presence of such absorption features, which would affect the observed transit depth according to our calculations (Fig.~\ref{HD209_trans_metals}). The residuals between the model and the broadband observations are about 6 and 20 sigma at the major Fe and Ca absorption signatures, respectively, while close to the calcium line the residuals relative to the medium resolution observations are more than three sigma. These results indicate that the abundances of Ca and Fe, at least, are decreased at the probed pressure levels of the HD 209458 b relative to the solar elemental abundances, a behavior anticipated due to condensation. Therefore we decided to consider atomic opacities of Na and K only for the remainder of this study. 

\section{Extended alkali chemistry}

The extended processes of alkali chemistry we consider in the model include the simulation of molecular chemistry, as well as of excited atomic state processes that could potentially reduce the abundances of Na and K. In addition, we investigate more aspects of their ion chemistry. An overview of the reactions involved and the rates used is provided in Tables~\ref{reac_grd} $\&$ \ref{reac_exc}.

\subsection{Molecular states} 
\begin{figure}[!t]
\centering
\includegraphics[scale=0.5]{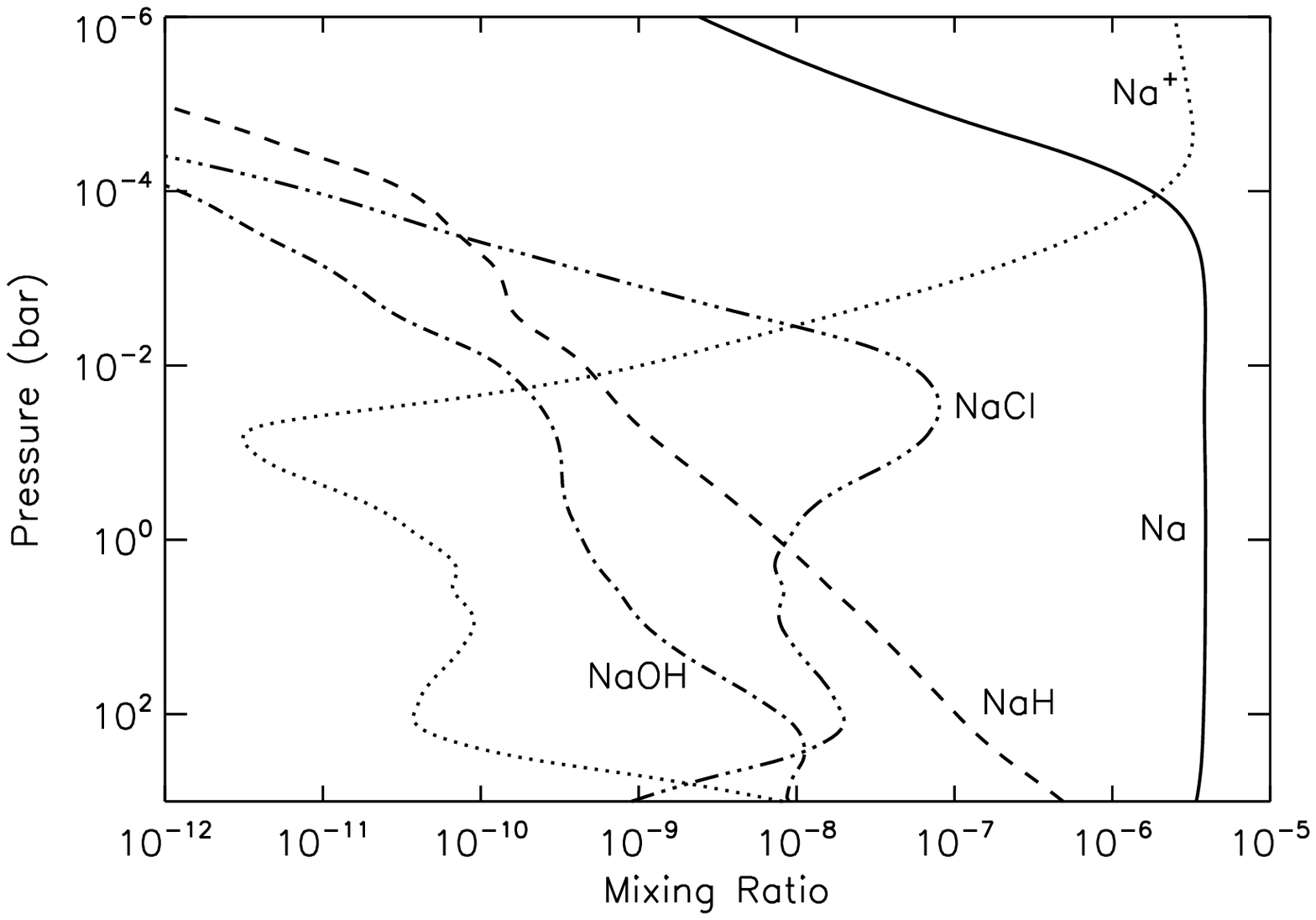}
\includegraphics[scale=0.5]{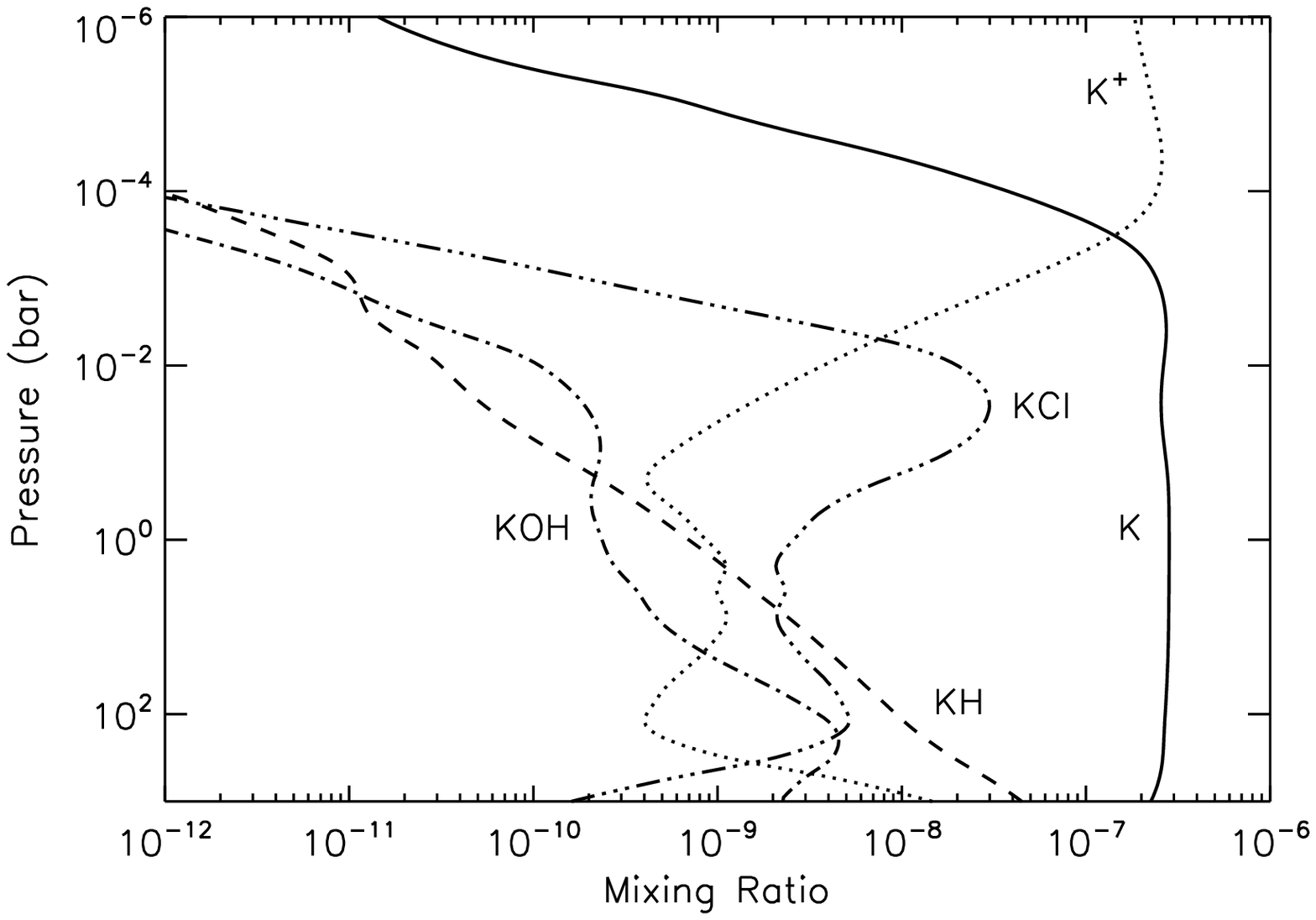}
\caption{Model results for the mixing ratios of different molecular structures of sodium (top) and potassium (bottom).}\label{molec}
\end{figure}

For the temperatures considered here, alkali atoms can form molecular structures. We follow the populations of NaH, NaOH, KOH, and KH, which are the thermochemically most abundant molecular structures at the thermal conditions assumed. We follow these populations kinetically so that we can identify any potential increases of their densities above the (otherwise low) thermochemical levels, due to quenching by photochemistry and dynamics. The chemical processes we consider for the description of these species are similar for Na and K (although with different rates for each element). For example, reaction of Na with water leads to NaOH: 
\begin{center}
Na + H$_2$O $\leftrightarrow$ NaOH + H, 
\end{center}
while reaction with hydrogen forms NaH:
\begin{center}
Na + H$_2$ $\leftrightarrow$ NaH + H. 
\end{center}
These molecules can thermally decompose back to sodium and the corresponding radicals:
\begin{center}
NaOH + M $\leftrightarrow$ Na + OH + M \\
NaH + M $\leftrightarrow$ Na + H + M, 
\end{center}
with both forward and reverse pathways important depending on the assumed temperature conditions. For the cases studied here the formation of these molecules is favored only deep in the atmosphere and their resulting abundance is always smaller than the atomic (Fig.~\ref{molec}). 

We also track the populations of NaCl and KCl, which require the inclusion of Cl and HCl in the study. Deep in the atmosphere chlorine is dominantly in the form of HCl in a balance with atomic Cl that is controlled through the reaction:
\begin{center}
HCl + H $\leftrightarrow$ Cl + H$_2$,
\end{center}
while at lower pressures reaction with alkali atoms leads to
\begin{center}
Na,K + HCl $\leftrightarrow$ NaCl, KCl + H
\end{center}
that affects the HCl population to a small degree.  We also consider the reaction
\begin{center}
HCl + OH $\leftrightarrow$ H$_2$O + Cl,
\end{center}
but found that it has a third role in the overall chlorine species profiles, after alkali atoms. Photolysis of HCl has a minor impact on its abundance and we included it with cross sections from \cite{Brion05}. The results show that close to the temperature minimum the abundances of both NaCl and KCl increase. KCl becomes large enough close to 10mbar that induces a small decrease in the atomic potassium abundance (Fig.~\ref{molec}). This decrease though is too small to reduce the transit depth of the potassium line to the level of the observations. Hence, molecular chemistry can not explain the low alkali abundances. Another conclusion we can derive from these results is that, notwithstanding the small differences in the profiles of the different species involved, Na and K have a very similar chemical behavior, as anticipated.

\newcounter{rctn}
\setcounter{rctn}{0}
\begin{table*}
\caption{Reactions of ground state alkali chemistry. The reaction constant provided corresponds to the reverse reactions. Units are in CGS.}\label{reac_grd}
\tiny{
\begin{tabular}{llllc}
\hline
& Reaction & Forward Rate & Reaction Constant &Notes/Reference \\
\hline
\addtocounter{rctn}{1} R\arabic{rctn}& Na + $h\nu$ $\rightarrow$ Na$^+$ + e$^{-}$ \\
\addtocounter{rctn}{1} R\arabic{rctn}& Na$^+$ + e$^-$ $\rightarrow$ Na + $h\nu$&k = 1.$\times$10$^{-11}$ $\exp(-50895/T)$& &\cite{Verner96}\\
\addtocounter{rctn}{2} R\arabic{rctn}& Na$^+$ + e$^-$ + M $\rightleftharpoons$ Na + M &k$_0$ = 3.43$\times$10$^{-14}$T$^{-3.77}$ &2.288$\times$10$^{+15}$T$^{1.506}$$\exp(-59630/$\rm{T}$)$&\cite{Su01}\\
 & & k$_{\infty}$ = 1.$\times$10$^{-7}$   & & Est\\

\addtocounter{rctn}{2} R\arabic{rctn}& NaH + H $\rightleftharpoons$ Na + H$_2$&k = 2.38$\times$10$^{-12}$T$^{0.69}$$\exp(-2360/T)$&1.160$\times$10$^{-01}$T$^{0.636}$$\exp(-29990/$\rm{T}$)$&\cite{Mayer66}\\
\addtocounter{rctn}{2} R\arabic{rctn}& Na + H$_2$O $\rightleftharpoons$ NaOH + H &k = 4.07$\times$10$^{-10}$ $\exp(-21900/T)$&3$\times$10$^3$T$^{-1.293}$$\exp(19210/$\rm{T}$)$&\cite{Jensen82}\\
\addtocounter{rctn}{2} R\arabic{rctn}& OH + Na + M $\rightleftharpoons$ NaOH + M &k$_0$ = 1.9$\times$10$^{-25}$T$^{-2.21}$$\exp(-41/$\rm{T}$)$&4.423$\times$10$^{+19}$T$^{1.115}$$\exp(-38360/$\rm{T}$)$&\cite{Patrick84}\\
 & & k$_{\infty}$ = 1.$\times$10$^{-11}$   & &Est \\
\addtocounter{rctn}{2} R\arabic{rctn}& H + Na + M $\rightleftharpoons$ NaH + M&k$_0$ = 1.9$\times$10$^{-25}$T$^{-2.21}$$\exp(-41/$\rm{T}$)$&4.873$\times$10$^{+19}$T$^{0.912}$$\exp(-20940/$\rm{T}$)$&Est as OH + Na + M\\
 & & k$_{\infty}$ = 1.$\times$10$^{-11}$   & & \\
\addtocounter{rctn}{2} R\arabic{rctn}& Na + H $\rightleftharpoons$ Na$^+$ + H$^-$&k = 1.$\times$10$^{-11}$ $\exp(-50895/T)$&2.81$\times$10$^{20}$T$^{-5.365}$$\exp(47650/$\rm{T}$)$&{Est}\\
\addtocounter{rctn}{2} R\arabic{rctn}& Na$^+$ + K $\rightleftharpoons$ K$^+$ + Na&k = 1.$\times$10$^{-11}$  &7.844$\times$10$^{-01}$T$^{0.032}$$\exp(-9244/$\rm{T}$)$&{Est}\\
\addtocounter{rctn}{1} R\arabic{rctn}& K + $h\nu$ $\rightarrow$ K$^+$ + e$^{-}$ \\
\addtocounter{rctn}{1} R\arabic{rctn}& K$^+$ + e$^-$ $\rightarrow$ K + $h\nu$&k$_0$ = 2.66$\times$10$^{-25}$T$^{-2.21}$$\exp(-48/$\rm{T}$)$& &\cite{Verner96}\\
\addtocounter{rctn}{2} R\arabic{rctn}& K$^+$ + e$^-$ + M $\rightleftharpoons$ K + M&k$_0$ = 3.43$\times$10$^{-14}$T$^{-3.77}$ &3.885$\times$10$^{+16}$T$^{1.155}$$\exp(-50680/$\rm{T}$)$&{Est as Na$^+$ + e$^{-}$}\\
 & & k$_{\infty}$ = 1.$\times$10$^{-7}$   & & \\

\addtocounter{rctn}{2} R\arabic{rctn}& KH + H $\rightleftharpoons$ K + H2&k = 2.38$\times$10$^{-12}$T$^{0.69}$$\exp(-2360/T)$&1.326$\times$10$^{-01}$T$^{0.678}$$\exp(-30370/$\rm{T}$)$&{Est as NaH}\\
\addtocounter{rctn}{2} R\arabic{rctn}& K + H$_2$O $\rightleftharpoons$ KOH + H &k = 5.$\times$10$^{-10}$ $\exp(-20000/T)$&1.789$\times$10$^{+03}$T$^{-1.277}$$\exp(16550/$\rm{T}$)$&\cite{Jensen79}\\
\addtocounter{rctn}{2} R\arabic{rctn}& OH + K + M $\rightleftharpoons$ KOH + M &k$_0$ = 2.66$\times$10$^{-25}$T$^{-2.21}$$\exp(-48/$\rm{T}$)$&8.414$\times$10$^{+18}$T$^{1.265}$$\exp(-40910/$\rm{T}$)$&\cite{Patrick84}\\
 & & k$_{\infty}$ = 1.$\times$10$^{-11}$   & & Est\\
\addtocounter{rctn}{2} R\arabic{rctn}& H + K + M $\rightleftharpoons$ KH + M&k$_0$ = 2.66$\times$10$^{-25}$T$^{-2.21}$$\exp(-48/$\rm{T}$)$&4.423$\times$10$^{+19}$T$^{0.866}$$\exp(-20570/$\rm{T}$)$&{Est as OH + K + M}\\
 & & k$_{\infty}$ = 1.$\times$10$^{-11}$   & & \\
 \addtocounter{rctn}{2} R\arabic{rctn}& HCl + $h\nu$ $\rightarrow$ H + Cl\\
\addtocounter{rctn}{2} R\arabic{rctn}& H + HCl $\rightleftharpoons$ H$_2$ + Cl &k = 2.4$\times$10$^{-11}$ $\exp(-1730/T)$&1.795T$^{-0.005}$$\exp(-478/$\rm{T}$)$&\cite{Allison96}\\
\addtocounter{rctn}{2} R\arabic{rctn}& HCl + OH $\rightleftharpoons$ H$_2$O + Cl &k = 6.84$\times$10$^{-19}$T$^{2.12}$$\exp(646/T)$&3.301$\times$10$^{+01}$T$^{-0.183}$$\exp(-8030/$\rm{T}$)$&\cite{Bryukov06}\\
\addtocounter{rctn}{2} R\arabic{rctn}& K + HCl $\rightleftharpoons$ KCl + H &k = 5.6$\times$10$^{-10}$ $\exp(-4170/T)$&2.707$\times$10$^{+01}$T$^{-0.648}$$\exp(710/$\rm{T}$)$&\cite{Husain88}\\
\addtocounter{rctn}{2} R\arabic{rctn}& Na + HCl $\rightleftharpoons$ NaCl + H &k = 4.$\times$10$^{-10}$ $\exp(-4090/T)$&3.825$\times$10$^{+01}$T$^{-0.646}$$\exp(2444/$\rm{T}$)$&\cite{Husain86}\\
\addtocounter{rctn}{2} R\arabic{rctn}& HCl + M $\rightleftharpoons$ H + Cl + M &k$_0$ = 7.31$\times$10$^{-11}$ $\exp(-41000/$\rm{T}$)$&6.4543320$\times$10$^{-19}$T$^{-1.632}$$\exp(50360/$\rm{T}$)$&\cite{Baulch81}\\
\hline
\end{tabular}}
\end{table*}

\begin{table*}
\caption{Reactions of excited state alkali chemistry. See Table~\ref{reac_grd} for details and text for discussion of the estimated rates.}\label{reac_exc}
\tiny{
\begin{tabular}{llllc}
\hline
& Reaction & Forward Rate & Reaction Constant &Notes/Reference \\
\hline
\addtocounter{rctn}{1} R\arabic{rctn}& Na[3P]  $\rightleftharpoons$ Na + $h\nu$&k = 6.15$\times$10$^{7}$  & &\cite{NIST}\\
\addtocounter{rctn}{1} R\arabic{rctn}& Na[4P] $\rightleftharpoons$ Na + $h\nu$&k = 2.74$\times$10$^{6}$  & &\cite{NIST}\\
\addtocounter{rctn}{1} R\arabic{rctn}& Na[3D] $\rightleftharpoons$ Na[3P] + $h\nu$&k = 3.43$\times$10$^{7}$  & &\cite{NIST}\\
\addtocounter{rctn}{1} R\arabic{rctn}& Na[4S] $\rightleftharpoons$ Na[3P] + $h\nu$&k = 1.32$\times$10$^{7}$  & &\cite{NIST}\\
\addtocounter{rctn}{1} R\arabic{rctn}& Na[4P] $\rightleftharpoons$ Na[4S] + $h\nu$&k = 6.63$\times$10$^{6}$  & &\cite{NIST}\\
\addtocounter{rctn}{1} R\arabic{rctn}& Na[4P] $\rightleftharpoons$ Na[3D] + $h\nu$&k = 1.05$\times$10$^{5}$  & &\cite{NIST}\\
\addtocounter{rctn}{2} R\arabic{rctn}& Na[3P] + M $\rightleftharpoons$ Na + M &k = 1.$\times$10$^{-11}$T$^{-0.50}$ &1. $\exp(-24239/$\rm{T}$)$&\cite{Earl74}\\
\addtocounter{rctn}{1} R\arabic{rctn}& Na[3P,4S,3D,4P] + $h\nu$ $\rightarrow$ Na$^{+}$ +  e$^{-}$ \\
\addtocounter{rctn}{2} R\arabic{rctn}& Na[3P] + H$_2$O $\rightleftharpoons$ NaOH + H &k = 2.2$\times$10$^{-6}$ $\exp(-21900/T)$&3$\times$10$^3$T$^{-1.293}$$\exp(-5119/$\rm{T}$)$&\cite{Muller80,Jensen82}\\
\addtocounter{rctn}{2} R\arabic{rctn}& Na[3P] + HCl $\rightleftharpoons$ NaCl + H&k = 1.6$\times$10$^{-12}$T$^{0.50}$ &3.825$\times$10$^{1}$T$^{-0.646}$$\exp(-21905/$\rm{T}$)$&{\cite{Weiss88}}\\
\addtocounter{rctn}{2} R\arabic{rctn}& Na[4S] + M $\rightleftharpoons$ Na + M &k = 9.$\times$10$^{-12}$T$^{-0.50}$&1. $\exp(-36777/$\rm{T}$)$&{Est as Na[4P]}\\
\addtocounter{rctn}{2} R\arabic{rctn}& Na[4S] + H$_2$ $\rightleftharpoons$ NaH + H &k = 9.$\times$10$^{-12}$T$^{-0.50}$ &8.62T$^{-0.636}$$\exp(-6777/$\rm{T}$)$&{Est as Na[4P]}\\
\addtocounter{rctn}{2} R\arabic{rctn}& Na[4S] + H$_2$O $\rightleftharpoons$ NaOH + H&k = 2.2$\times$10$^{-6}$ $\exp(-21900/T)$&3$\times$10$^3$T$^{-1.293}$$\exp(-17657/$\rm{T}$)$&{Est as Na[3P]}\\
\addtocounter{rctn}{2} R\arabic{rctn}& Na[3D] + M $\rightleftharpoons$ Na + M &k = 9.$\times$10$^{-12}$T$^{-0.50}$ &1. $\exp(-41682/$\rm{T}$)$&{Est as Na[4P]}\\
\addtocounter{rctn}{2} R\arabic{rctn}& Na[3D] + H$_2$ $\rightleftharpoons$ NaH + H &k = 9.$\times$10$^{-12}$T$^{0.50}$ &8.62T$^{-0.636}$$\exp(-11682/$\rm{T}$)$&{Est as Na[4P]}\\
\addtocounter{rctn}{2} R\arabic{rctn}& Na[3D] + H$_2$O $\rightleftharpoons$ NaOH + H&k = 2.2$\times$10$^{-6}$ $\exp(-21900/T)$&3$\times$10$^3$T$^{-1.293}$$\exp(-22562/$\rm{T}$)$&{Est as Na[3P]}\\
\addtocounter{rctn}{2} R\arabic{rctn}& Na[4P] + M $\rightleftharpoons$ Na[3D] + M &k = 9.$\times$10$^{-12}$T$^{-0.50}$ &1. $\exp(-1567/$\rm{T}$)$&\cite{Kleiber93}\\
\addtocounter{rctn}{2} R\arabic{rctn}& Na[4P] + H$_2$ $\rightleftharpoons$ NaH + H &k = 9.$\times$10$^{-12}$T$^{0.50}$ &8.62T$^{-0.636}$$\exp(-13249/$\rm{T}$)$&\cite{Kleiber93}\\
\addtocounter{rctn}{2} R\arabic{rctn}& Na[4P] + H$_2$O $\rightleftharpoons$ NaOH + H&k = 2.2$\times$10$^{-6}$ $\exp(-21900/T)$&3$\times$10$^3$T$^{-1.293}$$\exp(-24129/$\rm{T}$)$&{Est as Na[3P]}\\
\addtocounter{rctn}{2} R\arabic{rctn}& Na[4P] + HCl $\rightleftharpoons$ NaCl + H&k = 4.$\times$10$^{-12}$T$^{0.50}$ &3.825$\times$10$^{1}$T$^{-0.646}$$\exp(-40805/$\rm{T}$)$&\cite{Weiss88}\\
\addtocounter{rctn}{1} R\arabic{rctn}& Na[3P] + Na[4P] $\rightarrow$ Na$^+$ + Na + e$^-$&k = 1.$\times$10$^{-9}$  & &\cite{Carre84}\\
\addtocounter{rctn}{1} R\arabic{rctn}& 2 Na[3P] $\rightarrow$ Na$_2^+$ + e$^-$&k = 2.15$\times$10$^{-13}$T$^{0.50}$ & &\cite{Geltman88} \\
\addtocounter{rctn}{1} R\arabic{rctn}& Na[3P] + Na[3D] $\rightarrow$ Na$_2^+$ + e$^-$&k = 4.7$\times$10$^{-12}$T$^{0.50}$ & &\cite{Babenko95}\\
\addtocounter{rctn}{1} R\arabic{rctn}& Na$_2^+$ + e$^-$ $\rightarrow$ Na + Na[3P]&k = 1.$\times$10$^{-8}$  & &{Est}\\
\addtocounter{rctn}{1} R\arabic{rctn}& Na[4P] + K[4P] $\rightarrow$ Na + K$^+$ + e$^-$&k = 1.$\times$10$^{-9}$  & &\cite{Carre84}\\
\addtocounter{rctn}{1} R\arabic{rctn}& Na[3P] + K[5P] $\rightarrow$ Na + K$^+$ + e$^-$&k = 1.$\times$10$^{-9}$  & &\cite{Carre84}\\
\addtocounter{rctn}{1} R\arabic{rctn}& Na[4P] + K[5P] $\rightarrow$ Na + K$^+$ + e$^-$&k = 1.$\times$10$^{-9}$  & &\cite{Carre84}\\
\addtocounter{rctn}{1} R\arabic{rctn}& 2 K[4P] $\rightarrow$ K$_2^+$ + e$^-$&k = 5.$\times$10$^{-13}$T$^{0.50}$ & &\cite{Klucharev77}\\
\addtocounter{rctn}{1} R\arabic{rctn}& 2 K[4P] $\rightarrow$ K + K[5P]&k = 2.$\times$10$^{-13}$T$^{0.50}$ & &\cite{Namiotka97}\\
\addtocounter{rctn}{1} R\arabic{rctn}& K[4P] + K[5P] $\rightarrow$ K + K$^+$ + e$^-$&k = 1.$\times$10$^{-9}$  & &\cite{Carre84}\\
\addtocounter{rctn}{1} R\arabic{rctn}& K$_2^+$ + e$^-$ $\rightarrow$ K + K[4P]&k = 1.$\times$10$^{-8}$  & &{Est}\\
\addtocounter{rctn}{1} R\arabic{rctn}& K[4P] $\rightleftharpoons$ K + $h\nu$&k = 3.8$\times$10$^{7}$  &&\cite{NIST}\\
\addtocounter{rctn}{1} R\arabic{rctn}& K[5P] $\rightleftharpoons$ K + $h\nu$&k = 1.1$\times$10$^{6}$  & &\cite{NIST}\\
\addtocounter{rctn}{1} R\arabic{rctn}& K[3D] $\rightleftharpoons$ K[4P] + $h\nu$&k = 1.74$\times$10$^{7}$  & &\cite{NIST}\\
\addtocounter{rctn}{1} R\arabic{rctn}& K[5S] $\rightleftharpoons$ K[4P] + $h\nu$&k = 1.18$\times$10$^{7}$  & &\cite{NIST}\\
\addtocounter{rctn}{1} R\arabic{rctn}& K[5P] $\rightleftharpoons$ K[5S] + $h\nu$&k = 4.6$\times$10$^{6}$  & &\cite{NIST}\\
\addtocounter{rctn}{1} R\arabic{rctn}& K[5P] $\rightleftharpoons$ K[3D] + $h\nu$&k = 1.$\times$10$^{6}$  &&\cite{NIST}\\
\addtocounter{rctn}{2} R\arabic{rctn}& K[4P] + M $\rightleftharpoons$ K + M  &k = 3.0$\times$10$^{-12}$T$^{-0.50}$ &1. $\exp(-18594/$\rm{T}$)$&\cite{Earl74}\\
\addtocounter{rctn}{2} R\arabic{rctn}& K[4P] + HCl $\rightleftharpoons$ KCl + H&k = 1.6$\times$10$^{-12}$T$^{0.50}$ &2.707$\times$10$^{+01}$T$^{-0.648}$$\exp(-30060/$\rm{T}$)$&\cite{Earl74}\\
\addtocounter{rctn}{2} R\arabic{rctn}& K[5S] + M $\rightleftharpoons$ K[4P] + M &k = 3.0$\times$10$^{-12}$T$^{-0.50}$ &1. $\exp(-11448/$\rm{T}$)$&\cite{Lin84}\\
\addtocounter{rctn}{2} R\arabic{rctn}& K[3D] + M $\rightleftharpoons$ K[4P] + M &k = 3.0$\times$10$^{-12}$T$^{-0.50}$ &1. $\exp(-12175/$\rm{T}$)$&\cite{Lin84}\\
\addtocounter{rctn}{2} R\arabic{rctn}& K[5P] + M $\rightleftharpoons$ K[5S] + M &k = 6.$\times$10$^{-12}$T$^{-0.50}$ &1. $\exp(-5264/$\rm{T}$)$&\cite{Lin84}\\
\addtocounter{rctn}{2} R\arabic{rctn}& K[5P] + M $\rightleftharpoons$ K[3D] + M &k = 6.$\times$10$^{-12}$T$^{-0.50}$ &1. $\exp(-4537/$\rm{T}$)$&\cite{Lin84}\\
\addtocounter{rctn}{2} R\arabic{rctn}& K[5P] + H$_2$ $\rightleftharpoons$ KH + H &k = 1.2$\times$10$^{-13}$T$^{0.50}$ &7.54T$^{-0.678}$$\exp(-4936/$\rm{T}$)$&\cite{Lin84}\\
\addtocounter{rctn}{2} R\arabic{rctn}& K[5P] + H$_2$O $\rightleftharpoons$ KOH + H &k = 1.$\times$10$^{-13}$$\exp(-20000/T)$ &1.789$\times$10$^{3}$T$^{-1.277}$$\exp(-18757/$\rm{T}$)$&{Est as Na[3P]+H$_2$O}\\
\addtocounter{rctn}{2} R\arabic{rctn}& K[5P] + HCl $\rightleftharpoons$ KCl + H&k = 4.$\times$10$^{-12}$T$^{0.50}$ &2.707$\times$10$^{1}$T$^{-0.648}$$\exp(-34596/$\rm{T}$)$&Est as Na[3P]\\
\addtocounter{rctn}{1} R\arabic{rctn}& K[4P,5S,3D,5P] + $h\nu$ $\rightarrow$ K$^{+}$ +  e$^{-}$ \\
\hline
\end{tabular}}
\end{table*}

\subsection{Excited states}

\begin{figure}
\centering
\includegraphics[scale=0.5]{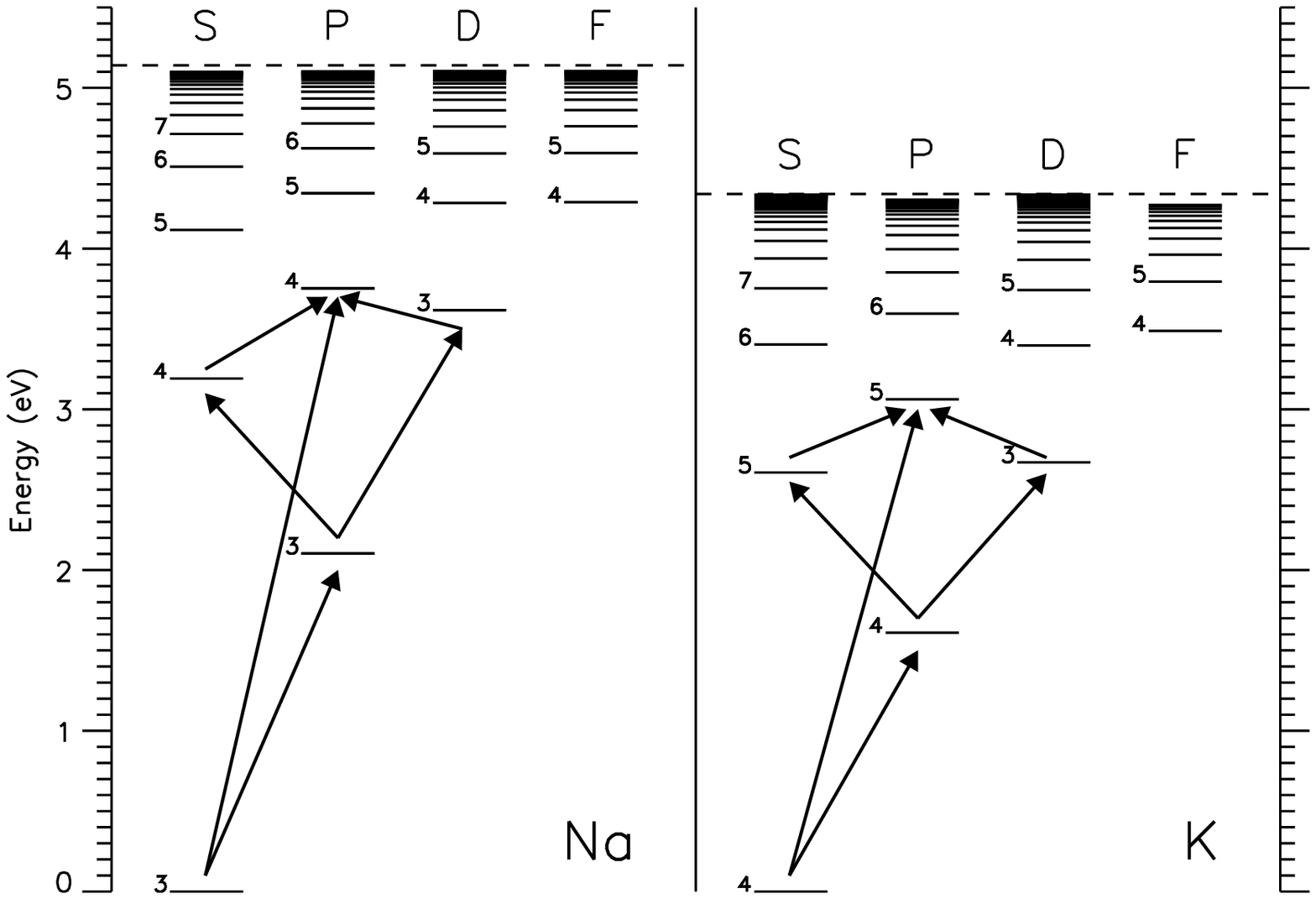}
\caption{Energy level diagram of the electronic states of Na and K. Arrows indicate the photon excitations considered in the calculations for the first four excited states of each atom. Dashed lines present the ionization limits.}\label{states}
\end{figure}

Due to  the large photon fluxes at visible wavelengths and the large oscillator strengths for the alkali resonant transitions, the photo-excitation rates of these states are orders of magnitude larger than the corresponding photo-ionization rates (Fig.~\ref{jrates}). Therefore a large population of excited alkali atoms can potentially exist in the exoplanet atmospheres. Processes forbidden or weak for ground state atoms could be efficient for the excited states and in this way drive the system further away from the thermochemical equilibrium. {\color{\clr}Moreover, if the resulting atomic state population of Na and K is much different from the LTE abundances, further modifications in the transit depth would be expected \citep{Barman02}.} Thus, a detailed description of these processes is necessary in order to evaluate their role in the abundances of alkali atoms. 

Two electronic states of Na and K can be efficiently excited from their grounds states directly. These are 3($^2$P) and 4($^2$P) for Na and 4($^2$P) and 5($^2$P) for K (Fig.~\ref{states}). For simplicity we reduce the nomenclature of these states to Na[3P], Na[4P], K[4P] and K[5P]. Each of these states is practically a doublet, but we are treating them as a single state in the photochemical calculations, i.e. we assume they have the same reactivity and radiative lifetimes. Although, different reactivities between the two terms have been observed for higher laying states, such effects are small and they will not affect our conclusions regarding the importance of the excited states chemistry on the overall abundance of the alkalis. Higher excited states of $^2$P symmetry can also be excited from the ground state, but the required shorter energy photons will be absorbed by other species higher up in the atmosphere and the oscillator strengths for these transitions decrease with increasing energy level. Further interaction of the first excited states with photons leads to the excitations of the [4S] and [3D] states of Na and the [5S] and [3D] states of K, which in their turn excite to the Na[4P] and K[5P] states, respectively (Fig.~\ref{states}). These are the four states we consider for each atom. We only consider excited states of the neutral alkali atoms, since excited states of the ions are separated by tens of eVs from the corresponding ground ion states.

{\color{\clr}We investigate in our calculations the different chemical reactivity of the excited states towards H$_2$, H$_2$O, and HCl. These reactions can lead to the formation of molecular complexes that could reduce the atomic abundances. To avoid interrupting the flow of the text we moved the details of this study to the appendix. Our results demonstrate that these processes can not significantly affect the atomic abundances of Na and K. In addition, the simulated non-LTE populations are not large enough to affect the transit depth observations. Our assumed atmospheric conditions and composition are different from those used by \cite{Barman02}, which is probably the reason for the different conclusions derived between the two studies about the efficiency of the non-LTE processes. Nevertheless, we note that the overall effect of the extended chemistry processes we include (see below) can affect the upper atmosphere abundances of the alkali atoms. }

 \subsection{Ion chemistry}

Excited alkali metals can further contribute to the ionization of the atmosphere. Direct photoionization from the radiation field 
\begin{center}
Na$^*$ + \it h\rm v $\rightarrow$ Na$^+$ + e$^-$,
\end{center}
is more efficient for the excited states since they are closer to the ionization potential, i.e. the ionization cross sections of the excited states extend to longer wavelengths than those of the ground state. Therefore their ionization rates can be significant since the stellar fluxes increase with increasing wavelength in this part of the spectrum. We calculated the photo-ionization cross section of the excited states following the analytical description from \cite{Zhang10} that is consistent with available measurements \citep{Wippel01, Amin05}.

Energy pooling reactions can also populate higher excited levels that can photo-ionize faster. For example, \cite{Huennekens83} have measured reaction rates for the process:
\begin{center}
2 Na$^*$ $\rightarrow$ Na[nl] + Na
\end{center}
and found cross sections of 23 and 16 \AA$^2$ for production of the [4D] and [5S] states, respectively, from collision of two Na[3P] atoms at 600 K. Theoretical calculations that are consistent with these measurements show that at higher temperatures the cross section for the formation of the [4S] state should increase to $\sim$40\AA$^2$ at 1000 K, while that for [5S] should decrease but less steeply \citep{Geltman89}. Energy pooling reaction for potassium where measured for collisions between [4P] states to produce [5P], [6S], and [4D] states \citep{Namiotka97}. According to our calculations, the loss rate of Na* states due to energy pooling is small compared to the other processes we consider and since we did not include any higher states than the Na[4P], we did not include this mechanism. For potassium we included the energy pooling rate to the K[5P] state, but it has a minor role. 

Other processes of direct ionization without a secondary photon involvement are those of Penning ionization:
\begin{center}
2 Na$^*$ $\rightarrow$ Na$^+$ + Na + e$^-$
\end{center}
and associative ionization:
\begin{center}
2 Na$^*$ $\rightarrow$ Na$_2^+$ + e$^-$
\end{center}
These mechanisms occur for both sodium and potassium, but also for combinations of the two. Associative ionization for Na has been observed for the [3P] states \citep{Huennekens83b} as well as from collisions between [3P] and [3D] states \citep{Babenko95}, while theoretical studies have demonstrated that the collision cross section for the associative ionization of two [3P] atoms at 1000 K is more than an order of magnitude smaller than the corresponding energy pooling cross section \citep{Geltman88}. Associative ionization for potassium was measured by \cite{Klucharev77} for two [4P] states, while \cite{Kimura81} measured the penning and associative ionization of excited Na atoms with ground state K atoms and found that production of K$^+$ and NaK$^+$ is possible but occurs only for sodium energy levels above the [6S] state. Thus, associative ionization can not have an important contribution to the ionization of the Na and K for the conditions we simulate, as demonstrated also by the resulting profiles of the dimer ions (Figs.~\ref{exct} $\&$ \ref{exctB}).


Penning ionization is possible for collisions between the two resonant states of sodium and potassium:
\begin{center}
Na[3P] + Na[4P] $\rightarrow$ Na + Na$^+$ + e$^-$ \\
K[4P] + K[5P] $\rightarrow$ K + K$^+$ + e$^-$,
\end{center} 
but also for the cross collision processes:
\begin{center}
Na[4P] + K[4P] $\rightarrow$ Na + K$^+$ + e$^-$ \\
Na[3P] + K[5P] $\rightarrow$ Na + K$^+$ + e$^-$ \\
Na[4P] + K[5P] $\rightarrow$ Na + K$^+$ + e$^-$ \\
\end{center}
These processes have large reaction cross sections and we used an estimated reaction rate of 10$^{-9}$ cm$^{3}$s$^{-1}$ based on the measurements for Na[3P] + Na[5S] collisions from \cite{Carre84}. Similar processes for the other states are also possible, but the populations of the resonant states are much larger and have the highest contribution to the alkali ionization through this pathway.

Inclusion of these processes does not affect the densities of Na and K below 1 mbar, and therefore does not change our transit depth results. Any change in the profiles occurs only in the upper atmosphere and could be only be observationally detected by looking at the cores of the Na and K lines that are probing this region of the atmosphere. Our calculations show that inclusion of the extended chemistry reduces the differential transit depth at the Na line by $\sim$15$\%$ at the smallest bands widths around the line cores.

\subsection{Further notes \& conclusions}

We finally add a few notes on the ground state ion chemistry. Radiative recombination of atomic ions usually leaves the neutral atom in an excited state from which it subsequently de-excites. We have not found any state specific information for the recombination of Na and K ions, thus we assumed that the produced neutrals are at the first excited states (Na[3P] and K[4P]). Assuming recombination to the Na[4P] and K[5P] states did not change significantly our results. In the 3-body recombination the excess energy is captured by the collision partner and the produced atoms should be on their ground state. We note that the released energy from the recombination of sodium (5.14 eV) is larger than the binding energy of the H$_2$ molecule (4.52 eV), thus the 3-body recombination could lead to the dissociation of H$_2$:
\begin{center}
Na$^+$ + e$^-$ + H$_2$ $\rightarrow$ Na + 2 H,
\end{center}
or at least to significant vibrational excitation. This process happens deep enough in the atmosphere for the atoms to rapidly recombine to H$_2$. Thus, the H$_2$ profile is not affected by the three body recombination. For potassium the ionization potential (4.32 eV) is less than (although comparable with) the H$_2$ binding energy and therefore the recombination will not dissociate the molecule although it will probably leave it in an excited vibrational level. 

For the high pressure limit of three body recombination we assumed a rate of 10$^{-7}$ cm$^3$s$^{-1}$ due to lack of information. Typically, recombination rates depend inversely on the square root of T, which means that the high pressure limit of the recombination rate close to the temperature minimum could be $\sim$50$\%$ smaller than deeper in the atmosphere. Our sensitivity test on this parameter show that we need to assume a high pressure limit of 10$^{-11}$ before the change in the total recombination rate becomes significant to affect the profiles of Na and K. Thus, modifications to the assumed recombination rate do not seem capable to reduce the alkali abundances. 

Due to their lower ionization potentials, charge transfer reactions of alkali atoms with other species are not efficient. Thus, the only charge transfer reactions to consider is: 
\begin{center}
\rm Na$^+$ + K $\leftrightarrow$ Na + K$^+$, 
\end{center}
with the forward rate in this case favored by 0.82 eV. Another ionization mechanism is that of the ion-pair formation:
\begin{center}
\rm Na + X $\rightarrow$ Na$^+$ + X$^-$ 
\end{center}
where X is a species (atom or molecule) with a large enough electron affinity to make this reaction efficient at high temperatures. The best candidates in our case (in terms of abundance and electron affinity) are H, O, and OH with electron affinities of 0.75, 1.46, and 1.83 eV, respectively, which means that the barriers for the ion-pairs formation will be 4.39, 3.68, and 3.31eV. The overall effect of these processes on the alkali atoms is small. It is interesting to note that Na and K also form stable negative ions with electron affinities of 0.55 and 0.50 eV, respectively. In this case the corresponding ion pair reactions should be limited among the alkali atoms:
\begin{center}
\rm Na + Na $\rightarrow$ Na$^+$ + Na$^-$ \\
\rm K + K $\rightarrow$ K$^+$ + K$^-$  \\
\rm Na + K $\rightarrow$ Na$^+$ + K$^-$  \\
\rm Na + K $\rightarrow$ Na$^-$ + K$^+$ 
\end{center}
The reduction in the involved energy barriers for these reactions is not as large as for the ion-pair formation processes with other species, and thus these mechanisms are not efficient. In any case, the ion-pair recombination (reverse of the above processes) is a barrierless reaction, therefore the net contribution of the ion-pair mechanism will be important only at very high temperatures and does not affect our conclusions for HD 209458 b. 

Associative ionization of sodium atoms with water molecules, although important in the atmospheres of our solar system, is not significant for the high temperature exoplanet atmosphere as the bond energy is small ($\sim$1eV) and the formed ligands would not survive in the atmosphere in significant abundances \citep{Gilligan01}. This effect is already demonstrated for the alkali dimers described above. {\color{\clr} The same conclusion also applies to the formation of clusters of alkali ions with H$_2$O or H$_2$, which can reduce the ion abundances due to their faster recombination rates. The high atmospheric temperatures would inhibit a significant population of the clusters, which even if they were possible would only make the agreement with the observations worse since the population of neutral atoms would increase. Therefore, we do not explicitly follow these processes in our calculations.}

Finally we investigated the role of the atmospheric mixing on the profiles of Na and K. Specifically we assumed a eddy mixing profile that is ten times smaller than the nominal profile we use. The resulting mixing ratios are not significantly different from the previous results in the region probed by the transit observations.  As a result, our attempts to match the Na differential transit depth and explain the non-detection of K with our photochemical model are not successful.

\section{The case of XO-2 \lowercase{b} }

Our calculations based on pure photochemical processes can not reproduce all available observations for the case of HD 209458 b. As we demonstrate in this section, however, our model can match the observations of both potassium and sodium absorption on XO-2 b (Sing et al., 2010, 2012).  This is interesting because it demonstrates that the atmosphere of XO-2 b may be fundamentally different from HD209458b where we cannot explain the STIS observations with our clear atmosphere model.



Apart from the physical parameters of this system (planet radius, mass, and distance from parent star), we need to re-evaluate the temperature profile and stellar flux in order to perform the photochemistry simulation. For the stellar spectrum we used results from the PHOENIX model \citep{Husser13} for a star with log(g) and effective temperature of XO-2 (www.exoplanet.eu) that provides the photospheric contribution ($\lambda$ $>$ 150 nm). We estimated the chromospheric contribution at shorter wavelengths from the HD 209458 stellar spectrum and from measurements of the ratios of FUV and EUV luminosity to the bolometric luminosity for XO-2 b, with updated values from \cite{Shkolnik13}. These measurements indicate that the chromospheric flux of XO-2 is weaker than that of HD 209458. Therefore, we performed calculations assuming chromospheric fluxes with the same structure as for the HD 209458 spectrum but reduced by factors of 2 and 4. Photons at chromospheric wavelengths are absorbed at high altitudes (see Fig.~\ref{RadField}) and based on our calculations any small modifications induced on the Na and K profiles are limited to altitudes above the 1 $\mu$bar level. Thus, variations in the chromospheric flux will not affect our conclusions for the deep-atmosphere alkali abundances. For the temperature profile, we note that the orbital distance of XO-2 b (0.03 au) leads to a similar skin temperature as for HD 209458 b (see dotted lines in Fig.~\ref{temps}). Therefore we assumed the same (disk) temperature profile for this calculation. 

\begin{figure}
\centering
\includegraphics[scale=0.5]{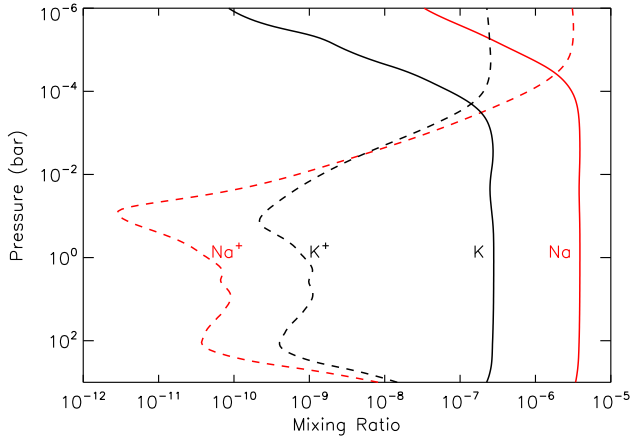}
\caption{Mixing ratio profiles of Na (red) and K (blue) in the atmosphere of XO-2 b. Solid lines present the neutral atoms and dashed the ions.}\label{XO2_comp}
\end{figure}

\begin{figure}[!t]
\centering
\includegraphics[scale=0.5]{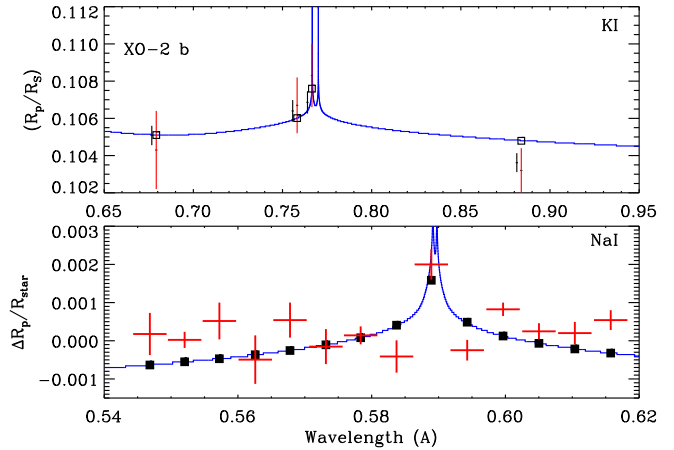}
\caption{Top: Planet size calculation at the potassium line for the exoplanet XO-2 b. The blue line is the high resolution calculation and the squares correspond to the R$_p$/R$_{star}$ for the observed wavelength bins. Vertical red and black lines show the observed planet size range at four wavelengths based on the \cite{Sing10} observations. The two sets of observations correspond to analysis based on the individual lines or combined analysis of all lines. Botttom: Comparison of the differential planet radius at the sodium core based on the calculations in blue, and the observations by \cite{Sing12} in red. The error bars in the measurements correspond to the 1 sigma level and the wavelength bins used have a width of 50 \AA. The black squares correspond to the model results at the resolution of the observations.}\label{xo2b}
\end{figure}

Our resulting Na and K profiles for this atmosphere are similar to the profiles we calculate for HD 209458 b (Fig.~\ref{XO2_comp}). The comparison with the transit observations for this case reveals a very good agreement with the R$_p$/R$_{star}$ measurements at the potassium line (Fig.~\ref{xo2b}). In addition our simulated broadband (4066-7590\AA) R$_p$/R$_{star}$ for the Na line is 0.1054 that is in excellent agreement with the measured value of 0.1052$\pm$0.0011 from \cite{Sing12}, while the differential planet radius around the sodium core is also consistent with the observations.

The contrast of the fit for XO-2 b with the calculations for HD 209458 b indicates that an explanation different from pure photochemical loss needs to be considered in order to explain the observations in the latter case. This explanation could be related to the presence of clouds in the atmosphere of HD 209458 b relative to a clear atmosphere in the case of XO-2 b, or a different elemental abundance of the alkali atoms.

In general, recent observations of different extrasolar planets point to variety of different scenarios.  Sometimes both Na and K are detected, sometimes only Na is detected and sometimes only K is detected (G. Ballester, personal communication).  In light of these observations we discuss the potential processes that could explain this variety below.  



\section{Discussion}

We discuss here possible scenarios that could help us interpret the Na and K observations in the atmosphere of HD 209458 b. These include modifications in the atmospheric temperature profile, the presence of clouds/haze, and the possibly different abundances of these elements relative to their solar values.

\subsection{Temperature effects}
One of the possible explanations proposed for the reduced Na transit depth and the lack of K is a sharp decrease in the temperature profile of HD 209458 b at pressures below $\sim$3 mbar \citep{Sing08b} that would reduce the abundance of both elements due to their capture into condensates (Na$_2$S) or molecular structures (KCl). Below this pressure level the temperature would have to remain high in order to reproduce the observed Na wing by \cite{Sing08a}. The vapor pressure curves for these conditions (see Fig.~\ref{temps}) indicate that temperatures lower by $\sim$300 K from the atmospheric skin temperature would be required for the alkali atoms to be significantly reduced. Such low temperatures however, are not supported by the observations.

At this point we need to make the distinction between day and night conditions. On one hand, the current circulation models for HD 209458 b from which our temperature profiles for day-time disk and limb conditions are taken, suggest that strong zonal winds efficiently redistribute the heat flux from the day side to the night side \citep{Showman09}. Thus, temperatures much lower than the skin temperature are not anticipated. On the other hand even if the night side is much colder than anticipated and the alkalis are captured in the form of condensates or molecules then the reduction in the observed transit depth would be only a factor of 2, since the observations probe both the dusk and dawn terminators. Thus, a reduced temperature scenario does not seem able to explain the observed transit depth. 

\begin{figure}[!t]
\centering
\includegraphics[scale=0.5]{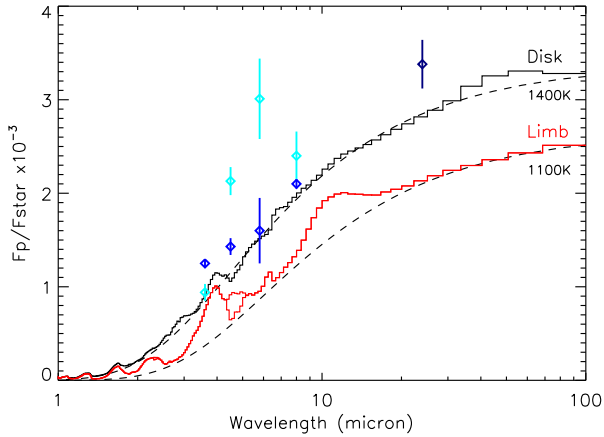}
\caption{Ratio of planet to star fluxes for HD 209458 b. The black and red solid lines present the emissions assuming the disk and limb temperature profiles. The dashed lines present the anticipated flux ratios assuming black bodies of 1400 and 1100 K, for the disk and limb respectively. Vertical bars with diamonds are the available observations from \cite{Knutson08} in cyan, \cite{Diamond14} in blue, and \cite{Crossfield12} in navy. The two red lines show the effect of CO addition close to 5 $\mu$m (higher opacity brings the model closer to the black body solution).}\label{thermal_emiss}
\end{figure}

An argument against globally lower temperatures arises from the thermal emission of HD 209458 b (Fig.~\ref{thermal_emiss}). With the assumed disk temperature profile and the calculated H$_2$O  and CO abundances from the photochemical model we derived a thermal emission spectrum that is consistent with the available observations from Spitzer. The latter include the initial observations by \cite{Knutson08} between 3.6 and 8 $\mu$m, and the more recent observations at these wavelengths by \cite{Diamond14} that suggest a lower flux ratio, as well as the 24 $\mu$m observations by \cite{Crossfield12}. The calculated spectrum indicates that a slightly warmer atmosphere would bring the thermal emission even closer to the observations. Hence, a colder atmosphere in order to the reduce the abundances of the alkali atoms, is not supported by these observations. 

\begin{figure}
\centering
\includegraphics[scale=0.5]{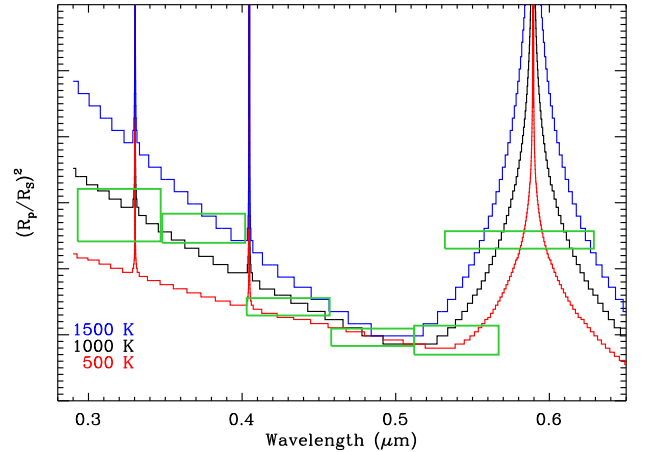}
\caption{Comparison of the H$_2$ Rayleigh slope of the simulated spectra at different temperatures with the slope of the HST/STIS observations (green boxes) from \cite{Deming13}.}\label{Compare}
\end{figure}

If we assume that the short wavelength slope in the transit depth observations is due to H$_2$ Rayleigh scattering \citep{desEtangs08}, we can derive an estimate of the atmospheric temperature. Assuming different isothermal conditions we can investigate the variation of the simulated slope of the transit depth at short wavelengths against the observations (Fig.~\ref{Compare}). Changing the atmospheric temperature, shifts the whole transit depth curve due to the changes in the atmospheric scale-height. Thus, in order to focus only on the variations of the Rayleigh slope we normalized our simulated transit spectra for different temperatures to 0.485 $\mu$m. For the comparison with the HST/STIS observations (which have small enough error bars to allow for this test) we scaled the simulated spectra so that they would result to the same transit depth at 0.485 $\mu$m. This comparison of the simulated slope with the observed slope demonstrates that temperatures close to 1000 K generate transit depth slopes that are in better agreement with the observed slope, compared to spectra generated for temperatures of 500 or 1500 K. A best fit search assuming only H$_2$ results in an isothermal profile of $\sim$1090 K, which is very close to the isothermal part of the limb temperature profile we use (Fig.~\ref{temps}). Note that this conclusion does not depend on which data set of the broad band observations we use \citep{Knutson07,Deming13}, since we only compare the relative absorption depths at different wavelengths that are approximately the same for the two studies. Therefore, if H$_2$ is responsible for the short wavelength slope of the observations, the temperature can not be low enough for Na and K to condense or be captured in molecular structures. 

\begin{figure}
\centering
\includegraphics[scale=0.5]{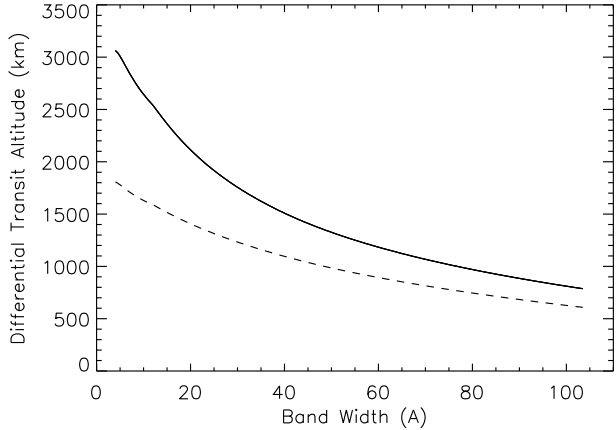}
\caption{{\color{\clr}Differential transit altitudes derived for a hypothetical  HD 209458 b atmosphere with a constant temperature of 1000 K, and under different assumptions for the Na abundance and vertical profile. The thick solid line is a composite of 3 cases of constant Na mixing ratio and abundances of 0.1, 1, and 10 times solar. The dashed line corresponds to the case of the Na profile derived by our photochemical model (see Fig.~\ref{Ions_mr}).}}\label{vidal}
\end{figure}

{\color{\clr}We note at this point that \cite{VidalMadjar11} derived from the differential transit depth observations a cold enough temperature profile for HD 209458 b to allow for the loss of Na due to condensation. Their retrieval was based on the assumption of a constant with altitude Na profile, though, which our calculations show is not realistic above $\sim$1 mbar (see Fig.~\ref{Ions_mr}). For a constant profile the temperature retrieval does not depend on the Na abundance, but this does not hold for a variable profile for which the differential transit depth has a different slope and magnitude (Fig.~\ref{vidal}). Thus, a more reliable temperature retrieval based on these observations needs to consider the variation of the Na profile with altitude due to the photochemistry. Given the apparent inconsistency between the mid- and low-resolution transit depth observations, we are reluctant to perform such an analysis before a better understanding of the observations is possible.}

\begin{figure}
\centering
\includegraphics[scale=0.5]{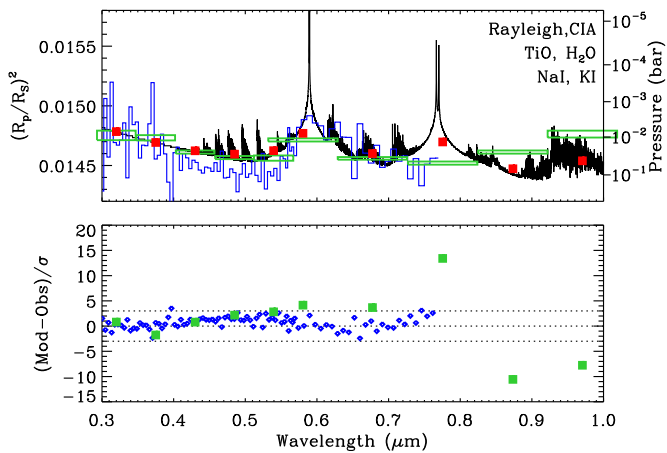}
\caption{Same as Fig.~\ref{HD209_trans} but with the inclusion of TiO, assuming a constant mixing ratio of 10$^{-10}$.}\label{TiO}
\end{figure}

A much warmer temperature scenario does not provide a solution either; even if more exotic metals become available in the atmosphere such as TiO or VO \citep{Fortney03, Desert08}, their spectral signature is not large enough to cover the broad wings of the potassium line. For example, assuming a mixing ratio of 10$^{-10}$ for TiO, the transit depth at the sodium line center is larger than the observations, while the K wings are not affected (Fig.~\ref{TiO}). Assuming a larger abundance (10$^{-9}$), the transit depth at wavelengths between 0.4 and 0.75 $\mu$m increases, but at shorter wavelengths it remains the same, as the TiO cross section sharply decreases for $\lambda < $ 0.4 $\mu$m. Thus, a large TiO abundance is inconsistent with the observed broadband slope, if the latter is due to H$_2$ Rayleigh scattering. If the slope is due to a different source of opacity, then a much higher temperature would be required in order to allow for large TiO abundance, enough to cover the K line and obscure the Na line. Such high temperatures would contradict the disk emissions, with which the temperature profile we use is consistent.

\subsection{Clouds and Hazes}
Another potential solution is that of a heterogenous opacity. In this scenario, the signatures of Na and K are reduced because a cloud or a photochemical haze is providing a spectrally homogeneous opacity that masks the alkali signature. As we demonstrated earlier this scenario, originally suggested as an explanation of the reduced differential transit depth around the Na doublet \citep{Charbonneau02}, is not consistent with the subsequent observations \citep{Ballester07,Knutson07,Sing08a}. On the other hand a reduced differential transit depth was retrieved in the subsequent observations from the high resolution spectra \citep{Sing08b} around the sodium core. One way to reconcile the two different conclusions derived from the same observations is to assume that the spectral region around the Na core is affected by another gaseous absorber with a relatively flat absorption spectrum. The problem with this scenario is that the Na doublet is one of the strongest absorption lines and in order to obscure its wings with another gaseous absorber, the abundance of the later would have to be too large. For example, TiO that has numerous lines around the Na core, would have to be present at a mixing ratio of 10$^{-8}$ before it substantially affects the differential transit depth. In that case numerous absorption lines at shorter wavelengths would have an observable signature in the transit spectrum, and generate a different spectral slope than observed. Therefore, it is clear that the analysis of the low and high resolution observations from HST/STIS provide inconsistent results. The same problem appears in the comparison of the low resolution analysis results \citep{Knutson07,Deming13} with the medium resolution spectra \citep{Sing08a}, with the later demonstrating a more pronounced Na wing compared to the former. All these differences demonstrate the difficulties inherent in the analysis of these observations. 

Nevertheless, if we momentarily neglect the differential transit observations (that are subject to higher uncertainty relative to the broadband results) a heterogeneous opacity can still be invoked to help understand the observations, provided that the haze or cloud structure can provide a short wavelength slope that is consistent with the observations. The probed pressures around the alkali lines indicate that this heterogenous opacity needs to be located close to the 10 mbar pressure level. It is important to keep in mind that we are observing the limb of the atmosphere. In this way the opacity of this cloud/haze does not have to be large in the vertical direction as long as the tangent opacity is consistent with the observations. Moreover, the consistency of the theoretical thermal emission spectrum with the observations (Fig.~\ref{thermal_emiss}) implies that heterogeneous opacity is small and/or that the particle size is much smaller than $\sim$1 $\mu$m in the probed region. Otherwise, scattering by the particles would modify the emitted radiation.

Heterogeneous opacities can also have an active photochemical role. For example, particles can accumulate free electrons in the atmosphere, and thereby decrease the recombination rates \citep[e.g.][]{Lavvas13}. This way the sodium and potassium abundances could be reduced deeper in the atmosphere. For the high temperatures of HD 209458 b, the large collision rates of the ions with the charged particles would limit the decrease in the recombination rate, though. Heterogenous reactions on the surfaces of the particles, however, could still affect the abundances of the alkali metals. All these processes require a dedicated study that will be presented in a future publication.

We should also note here that the case of heterogeneous opacity could also be consistent with the cases where only potassium is observed. In this case a low pressure opacity would be required (i.e. a photochemical haze rather than a cloud), in order to mask the Na line and with the appropriate slope in order to allow for the K line to be detected. The magnitude of the required slope could constrain the particle properties and be used to evaluate the feasibility of the haze scenario against other options. Moreover, the presence or not of line wings in the observations is a limiting factor for these scenarios, therefore high resolution observations, able to resolve these features are a requirement for future measurements.

\subsection{Elemental abundances}

\begin{figure*}[!t]
\centering
\includegraphics[scale=0.75]{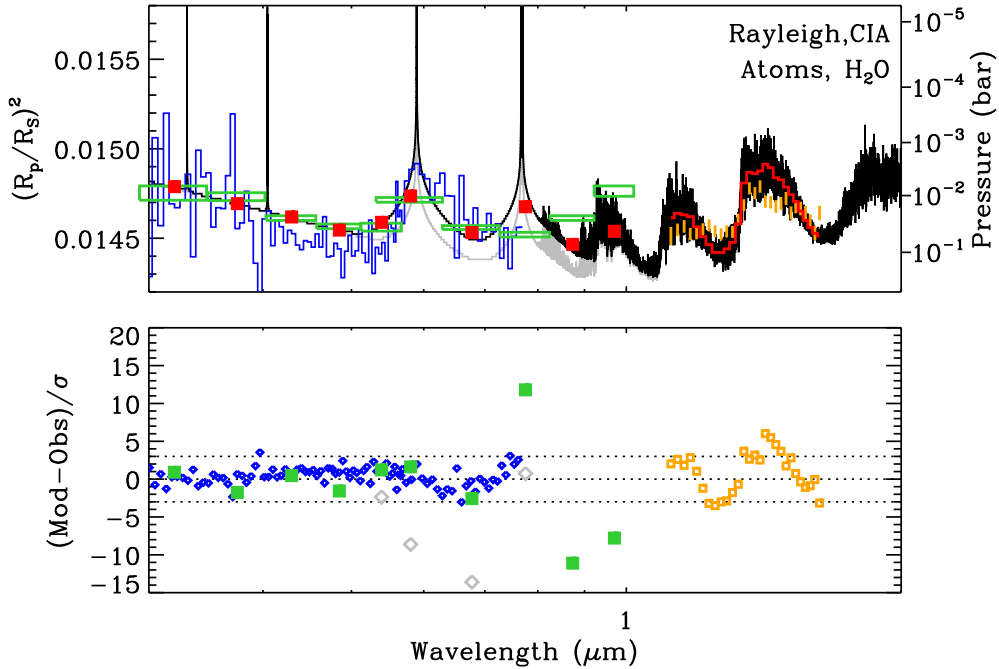}
\caption{The same as Fig.~\ref{HD209_trans} but zoomed in the region between 0.3 and 2 $\mu$m. The gray line in the top pager shows our model results in the case we divide the Na and K abundances by a factor of 10, while the gray diamonds in the lower panel show how the model-observation differences for this case. }\label{HD209_transit}
\end{figure*}

Up to now we assumed solar elemental abundances in all of our calculations. The simplest method to reduce the Na and K transit depths is to assume that both elements have much different abundances in the atmosphere of HD 209458 b. This difference can be absolute or relative. If the ratio of the two elements is the same as solar but the absolute abundance is smaller, then we would have to reduce the abundances by a factor of 10 relative to the solar value in order to make the transit depth at the potassium line consistent with the observations (Fig.~\ref{HD209_transit}). In this case though, the corresponding reduction of the Na abundance would make the sodium wings much more narrow than what is suggested by the STIS observations analyzed by \cite{Sing08b}, and the overall transit depth would be much lower than the broadband absorption retrieved by \cite{Knutson07} or \cite{Deming13}. Thus, a homogeneous decrease of the elemental abundances relative to the solar values is not consistent with the observations. 

Is there though a reason to anticipate that the Na/K ratio in HD 209458 b is similar to the solar ratio? Before answering this question we need to evaluate the degree at which we need to modify the assumed elemental abundances to bring the simulations in agreement with the observations. As discussed above, the solar abundance of K needs to be reduced by a factor of 10 to make its signature consistent with the transit depth spectrum from the \cite{Deming13}. Explaining the mid resolution HST/STIS observations, as presented in \cite{Desert08}, requires decreasing the solar K abundance by a factor of $\sim$50 to bring the transit depth at the line core \cite[the bin including the K line center in][]{Desert08} within 3 sigma of the observations. On the other hand, the solar Na abundance provides an acceptable fit to the sodium wing transit depth derived from the HST/STIS observations (neglecting the differential transit depth), implying that no major change is required for this element. These results imply that Na/K $\simeq$150-750 in the atmosphere of HD 209458 b, which is a substantial modification to the corresponding solar ratio, (Na/K)$_{\odot} \simeq$15. Thus, we can reformulate the above question to: \it is such a large abundance ratio realistic? \rm

There are two approaches to answer this question. The first involves the actual spectrum of HD 209458, which demonstrates the presence of both Na and K as seen from ground based observations \citep[e.g.][]{Jensen11}. Although an evaluation of the abundance of these elements in the star's atmosphere is not available, the structure and magnitude of the spectral lines of the two elements imply that the abundance of K is not two orders of magnitude smaller than that of Na. Although not conclusive, this is the only constraint available for this system currently.

From a wider perspective we can address the question of the relative abundances of Na and K by observing the statistics of their ratio in different stars. Observations of very low metallicity massive stars, which are representative of the first stars formed in the universe, can provide us with information about the initial elemental abundance of heavy elements in the interstellar medium. Measurements of the sodium and potassium abundance ratios relative to magnesium in such stars of our galaxy give mean values of [Na/Mg]=-0.8 dex and [K/Mg]=-0.3 dex, with scatter of 0.14 and 0.09, respectively \citep{Andrievsky10}. Although these values suggest lower elemental abundances of the alkalis relative to the solar values (as anticipated for the early stages of the universe) they demonstrate that the Na/K ratio ranges between 0.16 and 0.62 relative to the solar ratio. Observations of stars with similar properties as HD 209458 b and the sun show that Na and K abundances are close to the solar values \citep{Shimansky03,Adibekyan12}, while investigations of high metallicity A stars show that the variability of their alkali abundance is consistent with Na/K $<$ 100 \citep{Takeda11}. Thus, a consistent picture arises from these observations that the large Na/K values required to explain the observations are not likely, at least in stellar atmospheres. This conclusion though concerns only the stellar atmosphere rather than the planet.

Planetary atmospheres do not have to be of the same elemental abundance as the parent stars, as we well know from the solar system. It is important to note though that even if the elemental abundances in the atmosphere of HD 209458 b have been modified after the formation of the planet (e.g. due to migration in the stellar system followed by condensation and/or chemical modification), then we would expect Na and K to undergo the same changes, since they have very similar chemical behavior, as demonstrated above. In other words, we would not expect a preferential decrease of the potassium abundance over that of sodium during the evolution of the planetary atmosphere due to chemical processes only. Thus, any potential changes in the Na/K ratio have to be traced back to the planet formation from the protoplanetary disk. 

If the Na/K ratio varied across the protoplanetary disk that gave rise to the planetary system, then modifications of the atmospheric ratio relative to the stellar value are possible. Variations of the abundances of different elements on the disk can occur as a result of turbulence, chemistry, and radiation pressure. Due to the large oscillator strengths of the resonant lines, Na and K are subject to strong radiative acceleration. This effect will be important only at disk regions that are optically thin and where the alkali metals are not ionized, and therefore will only affect a small part of the disk's mass. Moreover, studies of this effect in debris disks demonstrate that, for low viscosity conditions, the Na/K ratio (at 100 AU) decreases relative to the solar value, contrary to the required modification in our case \citep{Xie13}. 

On the other hand, models coupling transport with photochemistry in protoplanetary disks (between 10 and 800 AU) demonstrate that the impact of turbulence varies among different gaseous species and ices, with the effect being more prominent on the latter component \citep{Semenov11}. This occurs as a result of the transport of ices/grains to regions of the disk where heterogeneous chemistry is more active (for the homogenous chemistry alone, no modifications were observed). The dominant condensates of alkalis in protoplanetary disks can be different from the condensates in planetary atmospheres, as the lack of cloud formation and stratification due to gravity, allows for the modification of condensates through interaction with other molecules. Thus, under thermochemical equilibrium Na and K should condense in the form of feldspar (NaAlSi$_3$O$_8$, KAlSi$_3$O$_8$) in protoplanetary disks compared to Na$_2$S and KCl in planetary atmospheres \citep{Lodders10b}. Due to the limited life time of protoplanetary disks ($<$10 Myr) and the variable lifetimes required for thermochemical equilibrium at different locations of the disk (due to variations in pressure and temperature) the abundance of the different components needs to be evaluated with detailed models that take into account the multiple processes involved. Unfortunately, a detailed study on the variation of Na and K over the protoplanetary disks is currently lacking. Any insight from observations is also limited. Spatially resolved spectra from the $\beta$-Pictoris debris disk demonstrate that Na is distributed over a large part for the disk, from 13 to 323 AU \citep{Brandeker04} but the corresponding distribution for K is unknown. 

Nevertheless we can evaluate the protoplanetary disk conditions required to explain the HD 209458 b observations. The necessary Na/K ratio retrieved from the observations implies that HD 209458 b should have emerged from a region of the disk that was deprived of K. This means that not only does K need to be lost from the gas phase but also its condensates or molecules need to be redistributed at a different radial location of the disk from where the planet is formed. Otherwise, gravitational heating (and the possible migration of the planet closer to the star) would have released the condensed alkalis back into the atmosphere\footnote{We expect that Na and K captured in ices/rocks would be differentiated in the same way in the planet interior and subject to similar evaporation mechanisms.}. At the same time Na would have to remain unaltered by the processes affecting the K abundance. 
The current understanding of giant planet formation suggests that HD 209458 b was formed further away from the star ($>$ 1 AU) and subsequently migrated to its current location. Hence, both Na and K should be present in condensed form at the planet formation region, and would be subject to similar differentiation processes. Therefore, Na and K condensates should not have very different responses to the differentiation mechanisms. Moreover any differentiation process will have to occur in a short timescale, before the disk is accreted to the star, in order to allow for the migration of the planet at its current short orbital distance. Hence it is difficult to explain the observations of HD209458b based on chemical and physical processes in the protoplanetary disk at the time of planet formation. 

Therefore, based on the expected primordial elemental abundances and alkali chemistry either in the protoplanetary disk or the atmosphere of the planet, HD209458b should not have a Na/K ratio that is substantially different from the solar ratio. While we cannot rule this possibility out unequivocally, it appears unlikely based on the above arguments.

\subsection{Observational issues}

Finally, there is one more aspect of the observations that requires some attention. The broad band observation on the red side of the K line demonstrate a significantly larger absorption than the model can produce. Absorption by H$_2$O would be consistent with these observations if the temperature was higher than the profile we assume, and in this case the simulated spectrum would be in better agreement with the Spitzer/IRAC observations at 5.8 and 8 $\mu$m. In this case though the model would provide too much absorption for the other observational constraints from HST/WFC and Spitzer. These differences among the different observations could potentially be related to differences in the analysis applied, but could also be related to temporal changes in the properties of the star. Similar effects could be possible at shorter wavelengths and could affect the relative levels of the alkali lines relative to the slope at shorter wavelengths. Simultaneous and repeated observations of the whole spectrum are required in order to clarify all these observational issues. 


\section{Conclusions}

Our photochemical calculations reveal that close-in exoplanets like HD 209458 b can have a significantly different ionospheric structure compared to the ionospheres of our solar system. From one hand, the proximity of the exoplanet to its parent star allows for very high ionization rates that result to large densities of ions and electrons (Fig.~\ref{Ion_den}). On the other hand the most striking feature is the vertical distribution of these ions. From our solar system examples, ionospheres are dominantly found in the region of the thermosphere as the high energy photons ($\lambda\le$1000 $\rm\AA$) required to ionize the typical molecules found there (H$_2$, N$_2$, O$_2$) can not penetrate at much deeper levels. The presence of heavier atoms with lower ionization potentials in the atmosphere of HD 209458 b, permits the ionization by longer wavelength photons ($\lambda\le$ 3000 $\rm\AA$), which are far more abundant. As a result, ionization rates for the heavy atomic species  are significantly higher and allow for large densities of ions and electrons in the lower part of the atmosphere (close to 0.1 mbar) compared to the corresponding densities generated due to thermal collisions. The potential implications of these high electron densities in the atmospheric properties are discussed in the accompanying paper \citep{Koskinen14}.

We have also performed a thorough study on the possible mechanisms that could explain the observations of Na and K in the atmosphere of HD 209458 b. In the past, the lack of a strong potassium signature along with the weak transit depth at the sodium core were attributed to either strong photo-ionization, a low temperature region in the atmosphere, the presence of clouds, or the different than solar elemental abundances of these elements in the atmosphere of HD 209458 b. We have demonstrated here that photochemistry alone can not explain the available observations, while we have also provided arguments against most of the other suggested scenarios. A much different temperature profile from that suggested by GCM models of this atmosphere are not supported by the observations, while a conveniently different elemental abundance is not likely based on observations of different stars, as well as of simulations of protoplanetary disks. The only case that appears more feasible is that of a heterogenous opacity from clouds or haze close to the 10 mbar pressure level that could screen the potassium absorption, but we showed that this scenario would not be consistent with the differential transit depth measurements around the Na core. Further studies on the role of clouds/aerosols on the heterogeneous chemistry of alkalis are required. On the other hand we have highlighted the inconsistencies among the different analyses of the same HST/STIS observations, which demonstrate the difficulties inherent in this process. Simultaneous and repeated observations of the whole spectrum (UV, visible and near IR) are required in order to resolve these inconsistencies and allow a better understanding for the abundance of alkali atoms in  exoplanet atmospheres. 

Notwithstanding the disagreement between the modeled and the implied by the observations abundance of K, we note that our results for the atmospheric electron density are not modified for pressures below $\sim$1 mbar, as this region is controlled by the photo-ionization of Na for which the model mixing ratio profile is in agreement with the observations.



\appendix

\section{Excited state chemistry}
The excited atoms can radiatively de-excite back to the ground state:
\begin{center}
\rm Na$^*$ $\rightarrow$ Na + \it h\rm v 
\end{center}
\cite[we use radiative de-excitation lifetimes from the NIST database, see][]{NIST}, collide with another species and de-excite:
\begin{center}
\rm Na$^*$ + M $\rightarrow$ Na + M, 
\end{center}
or partake in a chemical reaction: 
\begin{center}
\rm Na$^*$ + H$_2$O $\rightarrow$ NaOH + H  \\
\rm Na$^*$ + H$_2$ $\rightarrow$ NaH + H. 
\end{center}
In the above we used Na$^*$ to describe any possible excited state of sodium, but we consider similar processes for potassium as well. The relative importance of each process is still under investigation and the available studies have demonstrated that the efficiency of each process does not depend only on the energy of the states involved, but also on the orbital configuration of the excited atoms \citep[see review by][]{Lin02}. In other words, Na and K can have a significantly different excited state chemistry.

The ro-vibrational distribution of the hydride molecules (NaH and KH) in these processes can provide important information about the reaction mechanisms involved. Theoretical calculations have identified two types of mechanisms, an insertion mechanism and a harpoon-type process. In the first case the excited atom attacks H$_2$ by inserting itself between the two hydrogen atoms, while in the second case the alkali atom attacks the side of H$_2$ in a co-linear mode, whereby the excited electron is inserted into the H$_2$ molecule forming temporarily a negative ion that attracts the remaining alkali ion. The insertion mechanism leads eventually to the production of rotationally excited products, while the harpoon method favors the vibrational excitation of the resulting hydride molecule. These theoretical considerations are consistent with the experimentally derived rovibrational properties of the nascent hydride molecules \citep{Bililign92a,Bililign92b}. Moreover they can explain the variation of the reaction properties for different states, as well as different alkali metals. For example, the insertion method is favorable for small atoms that suffer a minimal repulsion from H$_2$ and is believed to describe the reaction mechanism for Na and Li, as well as the low laying states of K. For larger atoms (or excited states that have a larger effective size) the harpoon method is more favorable \citep[see][and references therein]{Chang08}. Although these theoretical calculation can provide us with a qualitative picture of the alkali chemistry, the current level of detail can not provide us with quantitative description of the involved rates required for our calculations. Therefore, estimates of these rates need to rely on the available laboratory measurements. 

The collisional deactivation rates for different excited states are usually reported in terms of a collision cross section, $\sigma$, between the excited atoms and the quenching species. The corresponding rate at temperature T can then be calculated from:
\begin{equation}
k = \sqrt{\frac{8k_BT}{\pi\mu}}\sigma
\end{equation}
where $\mu$ is the reduced mass of the colliding pair. For Na and K collisions with H$_2$, this reduces too:
\begin{equation}
k[cm^3s^{-1}] = 10^{-12}\sqrt{T}\sigma
\end{equation}
when $\sigma$ is given in $\rm\AA^2$. 

The collisional de-excitation of Na[3P] atoms by H$_2$ was measured by \cite{Earl74} at high temperatures (873K) who found a deactivation cross section of 10 $\rm\AA^2$. The formation of NaH from Na[3P] has a barrier 0.4-0.5 eV \citep{Motzkus97} and non-reactive quenching was found to dominate through:
\begin{center}
Na[3P] + H$_2$($\nu$=0) $\rightarrow$ Na + H$_2$($\nu\ge$1).
\end{center}
Na[3P] has enough energy to populate the first 3 vibrational levels of H$_2$ and a subsequent reaction of the vibrationally excited hydrogen with other Na[3P] atoms can form NaH:
\begin{center}
Na[3P] +  H$_2$($\nu\ge$1) $\rightarrow$ NaH + H,
\end{center}
with a rate of 1.1$\times$10$^{-9}$ cm$^{-3}$s$^{-1}$ \citep{Motzkus98}.
Deactivation of Na[4P] state to the Na[3D] state was found to proceed with a cross section of 7 $\rm\AA^2$ at 700 K \citep{Astruc88}, and deactivation of Na[4S] was observed to dominantly quench to Na[3P] with a cross section of 41.4 \AA$^2$ \citep{Astruc86}. In later studies, \cite{Kleiber93} found that the reactive and non-reactive quenching channels in the collision of Na[4P] with H$_2$ (at 470 K) proceed with a ratio of $\sim$1 and a total cross sections of $\sim$18 \AA $^2$. Moreover, \cite{Motzkus97} found that direct deactivation of Na[4P] to the ground state accompanied with the vibrational excitation of H$_2$ is unlikely, and instead non-reactive quenching of Na[4P] proceeds to the Na[3D] state with the excess energy transformed to translation energy of each colliding member. Therefore, although both Na[3P] and Na[4P] eventually form NaH, the mechanisms involved are different \citep{Motzkus98}. Based on these studies, we assume in our calculations non-reactive quenching of the Na[3P] state, and equal yields of reactive and non-reactive paths for the Na[4P] state. For the other sodium states we assume the same processes/rates as for the Na[4P] state, but we take into account the energy of each state in the calculation of the reverse reaction rates. 


The collisional de-excitation of different states of potassium by H$_2$ has been measured by \cite{Earl74} who derived total cross sections of 3 and 12 $\rm\AA^2$ for K[4P] and K[5P], respectively. Among the states we consider only K[5P] has enough energy for the chemical reaction. Early studies on the reactivity of this state have found that the relative contribution of the chemical pathway appears to be small (upper limit of 1$\%$ of total deactivation), while its quenching proceeds to the lower laying K[5S] and K[3D] states instead of the ground state K[4S] \citep{Lin84}. Later studies where able to detect the formation of KH from K[5P] confirming the presence of the reaction channel \citep{Liu96}, as well as the non-reactive de-excitation pathways \citep{Wong99}. Therefore, we assumed non-reactive quenching for the K[4P] state and a low yield (1$\%$) for reaction of the K[5P] state. For the other potassium states considered we assume the same processes/rates as for the K[4P] state. 

\begin{figure}[!t]
\centering
\includegraphics[scale=0.5]{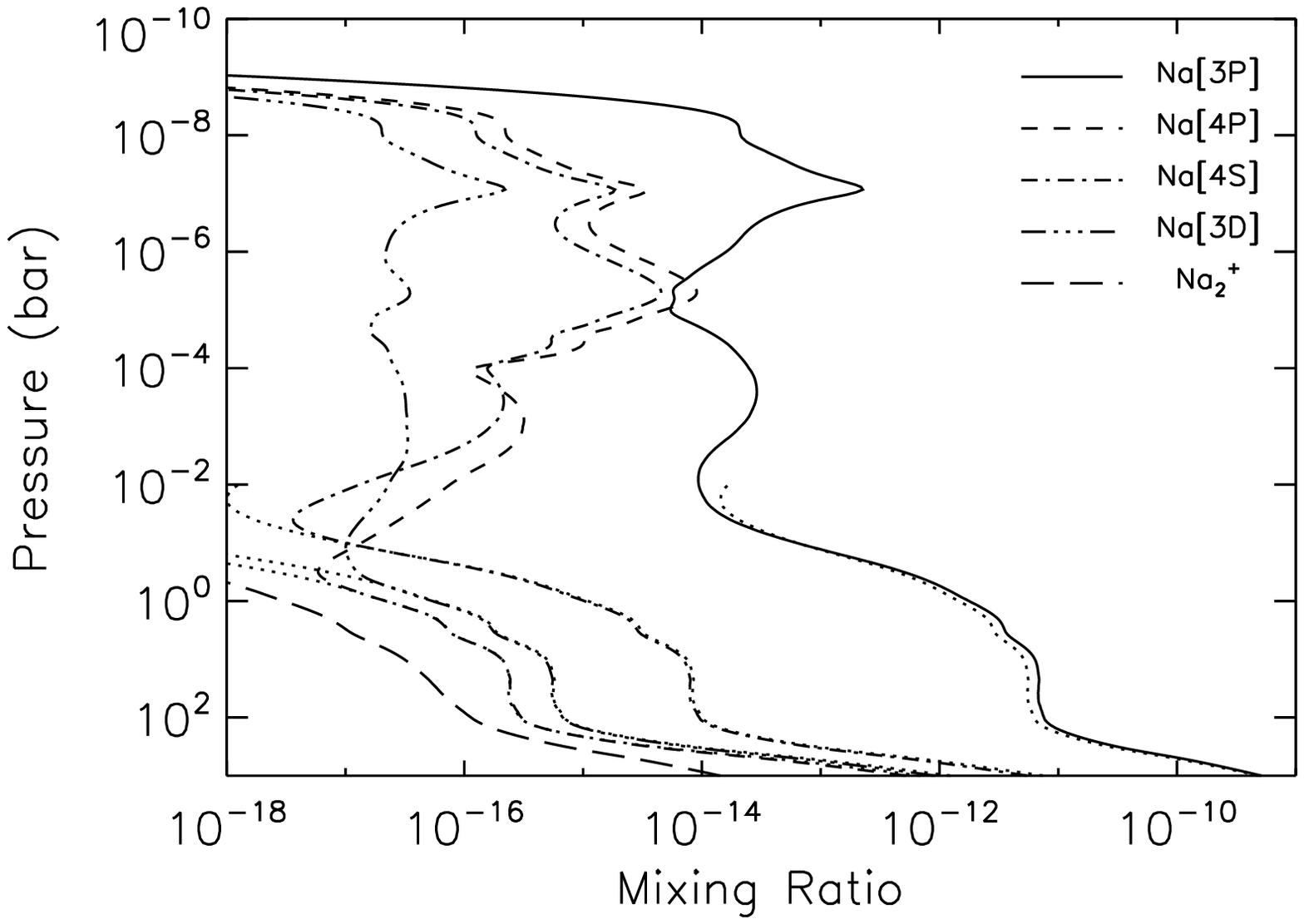}
\includegraphics[scale=0.5]{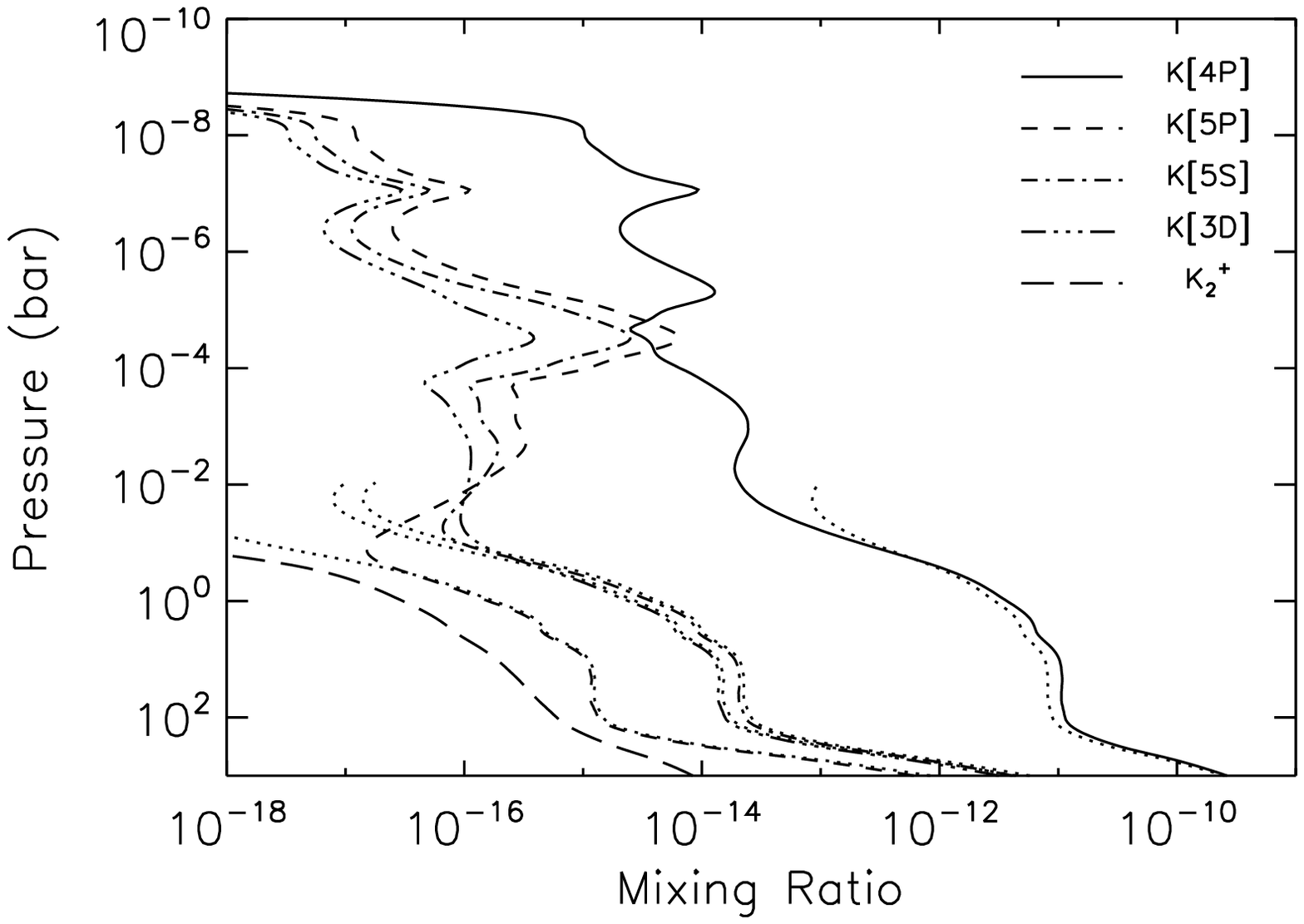}
\caption{Mixing ratios of different excited states of Na and K. The dotted lines present the corresponding mixing ration assuming LTE conditions.}\label{exct}
\end{figure}

\begin{figure}[!t]
\centering
\includegraphics[scale=0.5]{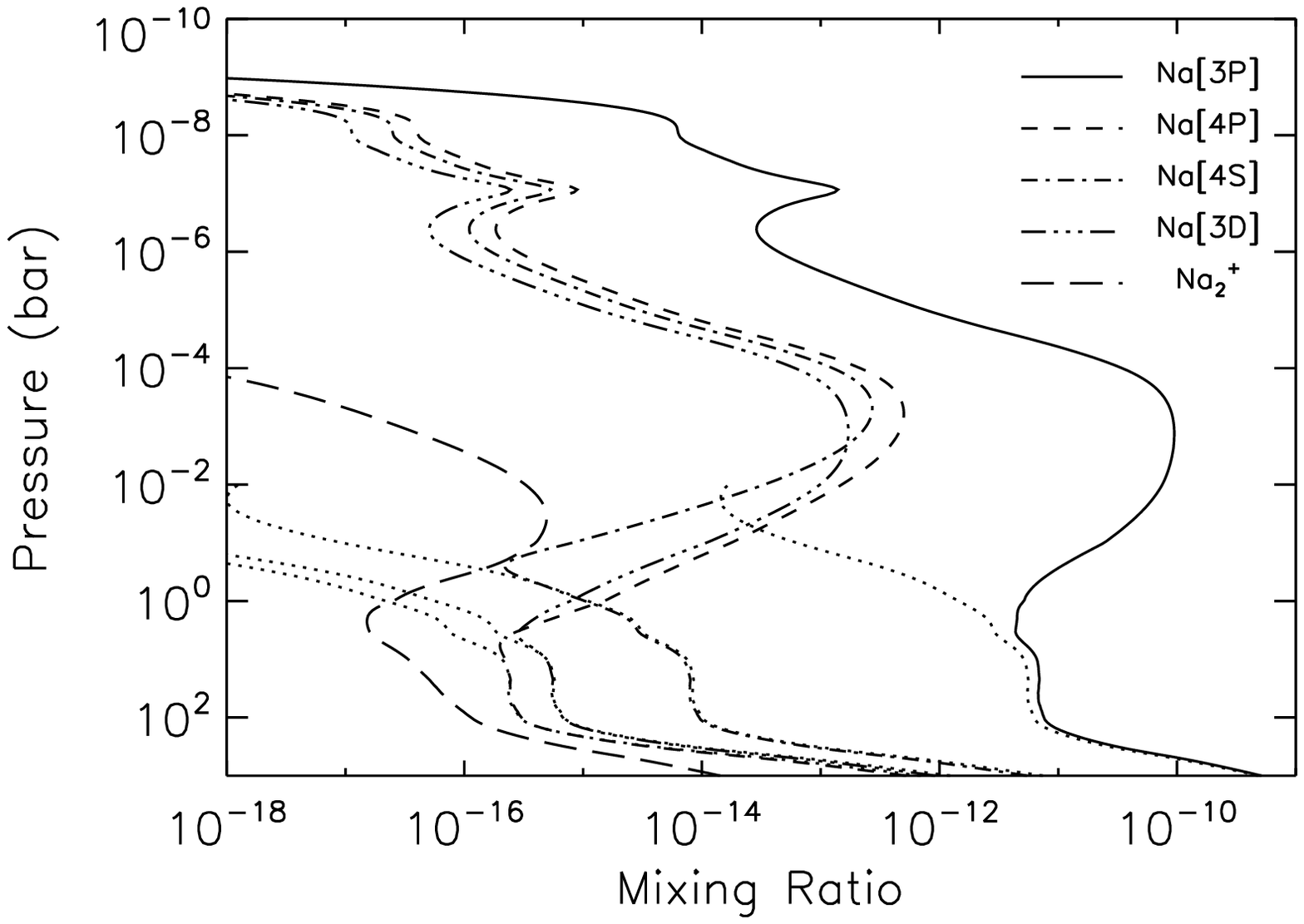}
\includegraphics[scale=0.5]{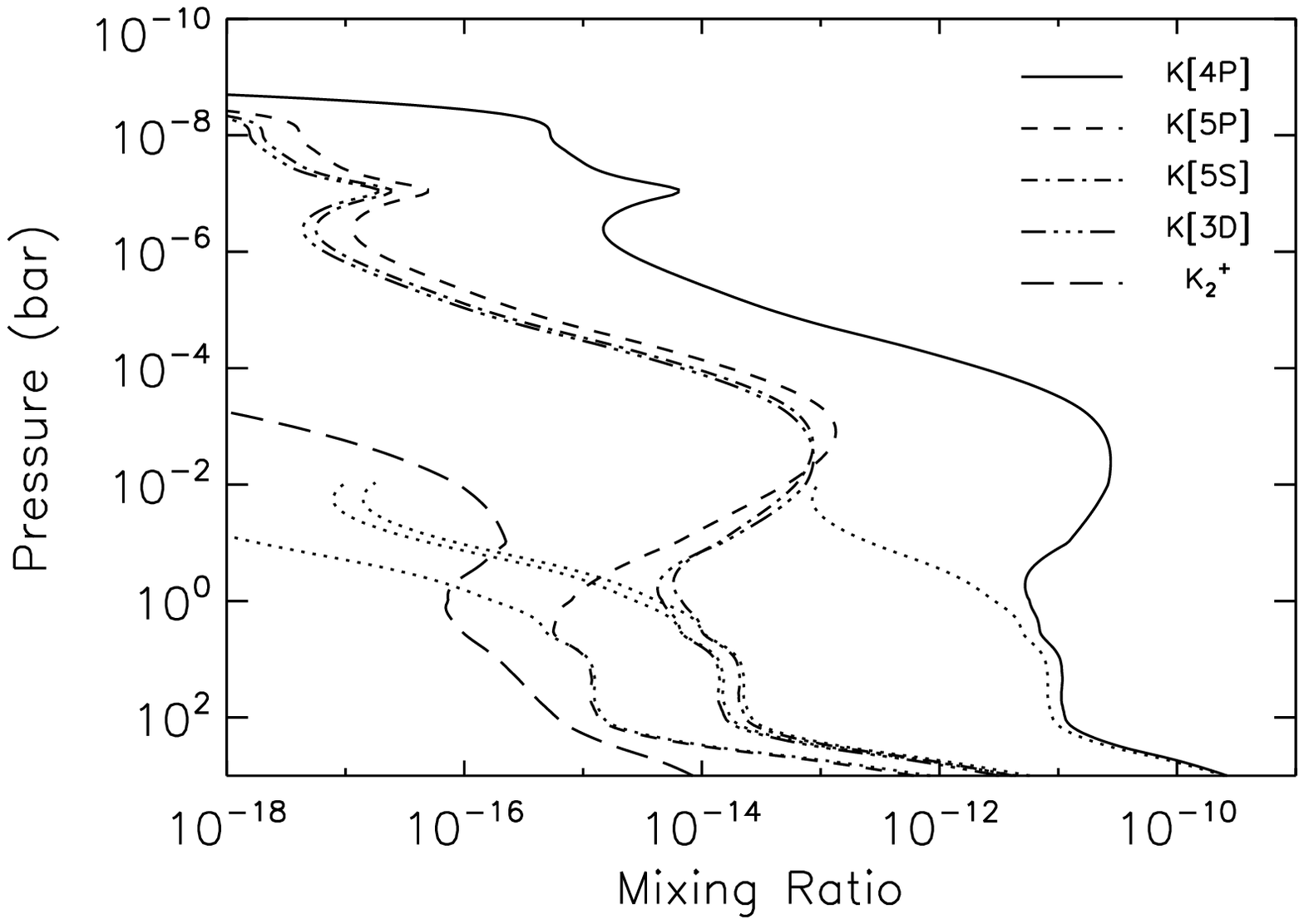}
\caption{Mixing ratios of different excited states of Na and K. The dotted lines present the corresponding mixing ration assuming LTE conditions. Here the attenuation of the stellar flux due to the resonance lines is not included.}\label{exctB}
\end{figure}

For reactions with H$_2$O, \cite{Earl74} measured a total deactivation cross section of 3 \AA $^2$ from the Na[3P] state, while \cite{Muller80} derived a ratio for the chemical reaction rate to form NaOH from Na[3P]/Na of 5350 at elevated temperature conditions (1925 K). Based on thermochemical consideration the reaction of H$_2$O with Na[3P] is exothermic with a reaction constant that is more than five orders of magnitude larger than the reaction constant for Na at $\sim$2000 K. Hence, the smaller value of the reaction ratio for the two states demonstrates that there is a reaction barrier involved. Reaction with higher energy states could have faster rates although we have not found any measurements for these processes. In lack of any information we have assumed the same rates as for the reaction of Na[3P], the latter estimated based on the observed ratio of Na[3P]/Na reaction rates and the measured ground state reaction from \cite{Jensen82}. For collisions of potassium with H$_2$O the corresponding total cross sections are 2 and 84 $\rm\AA^2$ for K[4P] and K[5P], respectively \citep{Earl74}, and due to lack of information we assume a chemical reaction yield with the same ratio of K[5P]/K rate as for sodium.

We have also considered the reactivity of the excited alkali atom states with {\color{\clr}HCl}. \cite{Weiss88} have shown that the reactivity to form NaCl, increases with increasing state energy for the states 3S, 3P, 5S and 4D. They measured an average cross section of $\sim$4 \AA$^2$ for the Na[3P] state, while for the other states we consider here we estimated a cross section of 10  \AA$^2$ for reaction based on their results. Due to the lack of any measurements for potassium we have assumed the same cross sections to form KCl as for the corresponding sodium states. 

The resulting profiles of excited states considered show that their populations are not large enough to affect the density of the alkali atoms  (Fig.~\ref{exct}). For these calculation{\color{\clr}s} we assumed that once photons are absorbed by alkali metals in the resonance lines, they are irreversible lost. In reality, when collisional de-excitation is not efficient and atoms have the time to radiatively de-excite, the photons are re-emitted. In order to the evaluate the role of these photons we also considered the extreme scenario that the stellar flux is not attenuated by the alkali resonant lines, i.e. the photons are forwardly re-emitted. Under this extreme scenario the populations of the excited states are drastically increased (Fig.~\ref{exctB}), but they still remain small enough not to affect the ground state populations of Na and K. Therefore excited state chemistry can not help resolve the disagreement of the model with the observations. 

It is interesting to note that in laboratory experiments of Na resonant excitation a complete ionization of the gas was observed \citep{Kushawaha82}. This effect was attributed to a two step process: first a free electron was produced through the associative (or Penning) ionization process, and then followed by the super-elastic scattering of the electron by excited sodium atoms that eventually provides enough energy to the electrons to ionize all the sodium gas \citep{Measures77}. In our case though the presence of H$_2$ will limit such effects as the excited alkali atoms will more easily transfer their energy to the vibrational levels of H$_2$. What happens to the excited hydrogen molecules, though? We evaluated the role of vibrationally excited H$_2$ in the formation of NaH through collisions with Na[3P] as described above. In this calculation we assumed that the vibrationally excited hydrogen is formed at the $\nu$=2 vibrational level from the reaction:
\begin{center}
Na[3P] + H$_2$($\nu$=0) $\leftrightarrow$ Na + H$_2$($\nu$=2)
\end{center}
The results show that for the conditions of the simulation NaH is not significantly affected by this process. 

Another contribution to the excitation of alkali atoms as well as H$_2$ could come from collisions with energetic electrons. In the region of the atmosphere we are interested the electron production is dominated by the photo-ionization of the alkali atoms. Collisions with H$_2$ will gradually reduce the energy of these electrons to the thermal energy, but in this process the molecules will be vibrationally excited. Collision of electrons with the alkali atoms is also possible, but the excitation rate of the different states from this process will be small compared to the direct excitation by photons, as demonstrated by the comparison of the photo excitation rates to the photo-ionization rates for each atom (Fig.~\ref{jrates}). The same argument also applies for the excitation of H$_2$, since direct excitation by collisions with Na[3P] would be more efficient in the lower part of the atmosphere. 

\section{Acknowledgments}
We acknowledge fruitful discussions with Caitlin Griffith, Gilda Ballester, and Travis Barman. We thank Steven Klippenstein for fruitful discussions on the role of ion clusters, Bertrand Plez for providing the TiO line list, and our reviewer for the suggested improvements on the manuscript. PL acknowledges support by the French Programme National de Plan\'etologie (PNP). TTK and RVY acknowledge support by the National Science Foundation (NSF) grant AST 1211514.

\bibliographystyle{apalike}
\bibliography{refs}

\end{document}